\documentclass[fleqn,usenatbib]{mnras}

\usepackage[T1]{fontenc}
\usepackage{ae,aecompl}

\usepackage{graphicx}	
\usepackage{amsmath}	
\usepackage{amssymb}	
\defcitealias{zha17a}{Primeval~I}
\defcitealias{zha17b}{Primeval~II}
\defcitealias{zha18a}{Primeval~III}
\defcitealias{zha18b}{Primeval~IV}
\defcitealias{zha19}{Primeval~V}

\title[Metal-poor degenerate brown dwarfs]{Primeval very low-mass stars and brown dwarfs -- VI. \\  Population properties of metal-poor degenerate brown dwarfs}

\author[Z. H. Zhang et al.]{Z. H. Zhang,$^{1}$\thanks{E-mail:
zenghuazhang@hotmail.com}\thanks{PSL Fellow} 
A. J. Burgasser,$^{2}$  M. C. G\'{a}lvez-Ortiz,$^{3}$ N. Lodieu,$^{4,5}$   
\newauthor 
M. R. Zapatero Osorio,$^{3}$  D. J. Pinfield$^{6}$ and  F. Allard$^{7}$   \\
$^{1}$GEPI, Observatoire de Paris, Universit{\'e} PSL, CNRS, 5 Place Jules Janssen, F-92190 Meudon, France \\
$^{2}$Center for Astrophysics and Space Science, University of California San Diego, La Jolla, CA 92093, USA \\
$^{3}$Centro de Astrobiolog{\'i}a (CSIC-INTA), Ctra. Ajalvir km 4, E-28850 Torrej{\'o}n de Ardoz, Madrid, Spain \\
$^{4}$Instituto de Astrof{\'i}sica de Canarias, E-38205 La Laguna, Tenerife, Spain \\
$^{5}$Universidad de La Laguna, Dept. Astrof{\'i}sica, E-38206 La Laguna, Tenerife, Spain \\
$^{6}$School of Physics, Astronomy and Mathematics, University of Hertfordshire, College Lane, Hatfield AL10 9AB, UK  \\
$^{7}$Univ Lyon, ENS de Lyon, Univ Lyon 1, CNRS, Centre de Recherche Astrophysique de Lyon UMR5574,  F-69007 Lyon, France 
}

\date{Accepted 2019 March 12. Received 2019 March 10; in original form 2019 February 12}

\pubyear{2019}

\begin{document}
\label{firstpage}
\pagerange{\pageref{firstpage}--\pageref{lastpage}}
\maketitle

\begin{abstract}
We presented 15 new T dwarfs that were selected from UKIRT Infrared Deep Sky Survey, Visible and Infrared Survey Telescope for Astronomy, and {\sl Wide-field Infrared Survey Explorer} surveys, and confirmed with optical to near infrared spectra obtained with the Very Large Telescope and the Gran Telescopio Canarias. One of these new T dwarfs is mildly metal-poor with slightly suppressed $K$-band flux. We presented a new X-shooter spectrum of a known benchmark sdT5.5 subdwarf, HIP 73786B. 
To better understand observational properties of brown dwarfs, we discussed transition zones (mass ranges) with low-rate hydrogen, lithium, and deuterium burning in brown dwarf population. The hydrogen burning transition zone is also the substellar transition zone that separates very low-mass stars, transitional, and degenerate brown dwarfs. Transitional brown dwarfs have been discussed in previous works of the Primeval series.
Degenerate brown dwarfs without hydrogen fusion are the majority of brown dwarfs. Metal-poor degenerate brown dwarfs of the Galactic thick disc and halo have become T5+ subdwarfs. We selected 41 T5+ subdwarfs from the literature by their suppressed $K$-band flux. We studied the spectral-type--colour correlations, spectral-type--absolute magnitude correlations, colour--colour plots, and HR diagrams of T5+ subdwarfs, in comparison to these of L--T dwarfs and L subdwarfs. We discussed the T5+ subdwarf discovery capability of deep sky surveys in the 2020s. 
\end{abstract}




\section{Introduction}
\label{sin}

Brown dwarfs (BDs) have insufficient mass to maintain their core temperature and pressure for sustained hydrogen fusion and are called `failed stars'. \citet[figs 4 and 5]{burr01} show that the 10 Gyr $T_{\rm eff}$/luminosity isochrones at a certain metallicity intersect with those at lower or higher metallicities. These intersection points between 10 Gyr mass--$T_{\rm eff}$ isochrones of different metallicities are used to define the steady hydrogen burning minimum mass (SHBMM) at the corresponding metallicity \citep[hereafter \citetalias{zha17b}]{zha17b}. 
Massive BDs with mass between $\sim$99 and 90 per cent of the SHBMM are in the substellar transition zone (STZ) and can reach states of minimum temperature and pressure to fuse hydrogen at low-rate in their cores to replenish the dissipation of their initial thermal energy and are referred to as transitional BDs \citep[T-BDs;][hereafter \citetalias{zha18a}]{zha18a}. 
The unsteady hydrogen fusion in T-BDs last for $\gg$ 10 Gyr \citep[fig. 6;][]{burr11}. 

Degenerate BDs (D-BDs) are the vast majority of BDs that do not fuse hydrogen ($\la$ 0.9 SHBMM) and keep cooling through their lifetime. D-BDs are important for studies of ultracool atmospheres and initial mass function. D-BDs could have spectral-types of M \citep{muen07,alle16}, L \citep{kir99,mart99}, T \citep{burg02}, and Y \citep{cus11,kirk11} types depending on their mass and age. Without energy supply from hydrogen fusion, D-BDs start their lives as warm late-type M dwarfs \citep[e.g.][]{rebo96}, then pass through the cool L dwarf domain sooner or later (depending on their mass), and will become T or Y dwarfs in time. The majority of known BDs discovered within 8 pc of the Sun are T dwarfs \citep{kir12}.  

Over 600 T dwarfs have been identified in the solar neighbourhood by modern sky surveys since the discovery of the first T dwarf, GL 229B \citep{naka95}. The spectral classification of T dwarfs has been widely discussed in the literature \citep[e.g.][]{burg02,bur03a,bur06,kir05}. T dwarfs emit most of their flux in the near-infrared (NIR), which are dominated by H$_2$O and CH$_4$ absorption bands with increasing strength towards later types. 

D-BDs with subsolar metallicity ([Fe/H] $\la -0.3$) in the thick disc usually are older than $\sim$8 Gyr \citep{kili17} and would have cooled to below $\sim$1000 K into mid-T to Y subdwarfs (\citetalias{zha18a}; \citealt[][hereafter \citetalias{zha18b}]{zha18b}). Since metal-poor T-BDs in the substellar transitional zone are stretched from $\sim$L3 to T4 types \citep{zha18c}. The number ratio between T and L subdwarfs in the solar neighbourhood should be much larger than that of  T and L dwarfs (\citealt{burg09}; \citetalias{zha17b}). However, metal-poor T dwarfs/subdwarfs are faint and rare, thus also difficult to identify. Large-scale infrared surveys such as the UKIRT Infrared Deep Sky Survey \citep[UKIDSS;][]{law07} and the {\sl Wide-field Infrared Survey Explorer} \citep[{\sl WISE};][]{wri10} provide good opportunities in finding T dwarfs.

Metal-poor T dwarfs or T subdwarfs have been reported in the literature \citep{burg02,burn10,burn14,murr11,pinf12,pinf14,mace13b}. Some of them are wide companions to metal-poor stars with known metallicities. These known T subdwarfs are mostly thick disc members according to their kinematics. The classification of T subdwarfs is based on direct comparison of their $J$- and $H$-band spectra to those of T dwarfs. T5+ subdwarfs have fainter $J$ and $H$ band absolute magnitudes than T dwarfs with the same spectral-types  \citep[fig. 22;][hereafter \citetalias{zha17a}]{zha17a}. 

T subdwarfs have higher $Y/J$ band flux ratio, suppressed $K$-band flux, and bluer $YJHK$ colours compared to T dwarfs \citep{bur06b,burn14}. The $K/J$ and $Y/J$ indices were used to probe the metallicity of T dwarfs \citep{burg02,bur06b}. \citet{mace13a} noted that $\sim$32 known late-type T dwarfs are bluer (with suppressed $K$-band flux) than T dwarf spectral standards. The esdT and usdT suclasses (halo T subdwarfs) with [Fe/H] $< -1.0$ have not yet been identified or defined. However, they are expected to have bluer $YJHK$ colours, higher $Y/J$ flux ratio, and flatter $K$-band flux than known T subdwarfs. 

This is the sixth paper in a series titled {\sl Primeval very low-mass stars and brown dwarfs}. The first to the fifth papers of the series are focused on L subdwarfs (\citetalias{zha17a}; \citetalias{zha17b}; \citetalias{zha18a}; \citetalias{zha18b}; \citealt[hereafter \citetalias{zha19}]{zha19}). In this paper, we present the population properties of metal-poor D-BDs. Section \ref{sob} presents a search for T dwarfs with wide sky surveys and spectroscopic follow-ups. Spectral classification of new T dwarfs is presented in Section \ref{ssc}. We discussed the physics and evolution of BDs in Section \ref{sebd}. Section \ref{spp} presents T subdwarf population properties based on 41 known objects in the literature. We discussed the discovery capability of future large-scale sky surveys in  Section \ref{sfss}. The summary and conclusions are presented in Section \ref{ssum}.

\begin{table*}
 \centering
  \caption[]{UKIDSS and {\sl WISE} photometry of our T dwarf candidates. }
\label{tdcand}
  \begin{tabular}{r c  c c c c c l l}
\hline
    Name~~~~~~~~~~~~ & $Y$	 & $J$ & $H$ & $K$ & $W1$ & $W2$ & SpT & SpT1 \\	
\hline
ULAS J000509.42+101407.7 & 19.35$\pm$0.08 & 18.15$\pm$0.05 & 17.72$\pm$0.05 & 17.79$\pm$0.11 & 16.93$\pm$0.11 & 16.23$\pm$0.22 & T3 & --- \\
ULAS J000844.34+012729.4 & 18.20$\pm$0.04 & 16.99$\pm$0.02 & 17.40$\pm$0.06 & 17.54$\pm$0.10 & 17.02$\pm$0.14 & 14.83$\pm$0.07 & T6.5$^a$ & T6.5$^b$  \\
ULAS J002136.00+155227.3 & 19.05$\pm$0.06 & 17.87$\pm$0.04 & 17.68$\pm$0.07 & 17.93$\pm$0.14 & 17.13$\pm$0.14 & 15.56$\pm$0.15 & T4p \\
ULAS J004030.41+091524.8 & 19.20$\pm$0.08 & 18.16$\pm$0.05 & 18.13$\pm$0.14 & 18.13$\pm$0.21 & 17.08$\pm$0.15 & 15.94$\pm$0.23 & T4.5  &---\\
ULAS J012252.58+043735.3 & 19.90$\pm$0.12 & 18.49$\pm$0.06 & 18.23$\pm$0.13 & 18.03$\pm$0.15 & 17.21$\pm$0.13 & 16.47$\pm$0.25 & T2.5  &---\\
VIK J012834.94$-$280302.6 & 18.84$\pm$0.03 & 17.71$\pm$0.01 & 17.73$\pm$0.04 & 17.97$\pm$0.07 & 18.07$\pm$0.28 & 15.35$\pm$0.10 & T5  &---\\
ULAS J012947.35+151143.1 & 19.50$\pm$0.16 & 18.23$\pm$0.06 & 18.04$\pm$0.09 & 17.88$\pm$0.16 & 17.70$\pm$0.20 & 16.09$\pm$0.19 & T4.5  &---\\
ULAS J014443.26+014741.0 & 19.15$\pm$0.08 & 17.94$\pm$0.05 & 18.24$\pm$0.13 & --- & 17.61$\pm$0.19 & 15.97$\pm$0.16 & T5  &---\\
ULAS J083346.88+001248.0 & 19.61$\pm$0.11 & 18.21$\pm$0.05 & 17.97$\pm$0.08 & 17.43$\pm$0.09 & 16.72$\pm$0.09 & 15.67$\pm$0.13 & T3  &---\\
VIK J091641.18$-$010911.5 & 18.89$\pm$0.03 & 17.75$\pm$0.02 & 18.05$\pm$0.05 & 18.18$\pm$0.08 & 18.24$\pm$0.38 & 15.78$\pm$0.15 & T6 \\
ULAS J105131.80+025751.0 & 18.48$\pm$0.04 & 17.46$\pm$0.03 & 17.77$\pm$0.10 & 17.94$\pm$0.16 & 16.97$\pm$0.13 & 15.09$\pm$0.09 & T7 & T6.5$^b$ \\
ULAS J112059.40+121904.4 & 19.20$\pm$0.07 & 17.93$\pm$0.04 & 18.00$\pm$0.05 & 17.94$\pm$0.12 & 17.88$\pm$0.26 & 15.89$\pm$0.17 & T4.5 & T5.5$^c$ \\
ULAS J124639.32+032314.2 & 18.95$\pm$0.06 & 17.60$\pm$0.03 & 18.05$\pm$0.13 & --- & 17.63$\pm$0.22 & 15.78$\pm$0.17 & T5 & T5$^c$ \\
VHS J132141.70$-$092446.2 & 17.97$\pm$0.02 & 16.71$\pm$0.01 & 16.69$\pm$0.02 & 16.77$\pm$0.05 & 16.36$\pm$0.07 & 14.92$\pm$0.07 & T4.5 &--- \\
ULAS J135950.14+062553.9 & 19.43$\pm$0.12 & 18.27$\pm$0.06 & 17.78$\pm$0.06 & 17.64$\pm$0.10 & 16.96$\pm$0.10 & 15.99$\pm$0.14 & T3 &--- \\
ULAS J150457.65+053800.8 & 17.64$\pm$0.02 & 16.59$\pm$0.02 & 17.05$\pm$0.04 & 17.41$\pm$0.09 & 16.00$\pm$0.05 & 14.23$\pm$0.04 & sdT5.5 & T6$^d$ \\
\hline
ULAS J092827.35+034658.5 & 19.29$\pm$0.08 & 18.09$\pm$0.05 & 18.29$\pm$0.15 & --- & 17.38$\pm$0.15 & 15.55$\pm$0.11 & cand$^e$ &--- \\
VHS J015058.31$-$015547.8 & 20.04$\pm$0.16 & 18.76$\pm$0.07 & 18.82$\pm$0.18 & --- & 17.10$\pm$0.12 & 16.17$\pm$0.19 & cand$^e$ &\\
ULAS J110446.63+032211.9 & 19.57$\pm$0.08 & 18.32$\pm$0.04 & 18.01$\pm$0.07 & 18.08$\pm$0.15 & 17.75$\pm$0.24 & 15.88$\pm$0.17 & cand$^e$ &--- \\
ULAS J021719.12+044239.8 & 19.29$\pm$0.08 & 18.09$\pm$0.06 & 17.83$\pm$0.07 & 18.15$\pm$0.18 & 17.21$\pm$0.12 & 15.88$\pm$0.14 & cand$^e$ &--- \\
ULAS J083316.82+031403.2 & 19.21$\pm$0.07 & 18.23$\pm$0.04 & 17.92$\pm$0.08 & 17.75$\pm$0.11 & 16.89$\pm$0.11 & 15.55$\pm$0.13 & cand$^e$ &---\\
\hline
\end{tabular}
\begin{list}{}{}
\item[Note.] SpT is adapted spectral-type, and SpT1 is photometric spectral-type in the literature. $^a$Candidate of unresolved BD binary of two T6+T8 dwarfs. $^b$Known T dwarf candidates identified by photometry \citep{skrz16}. $^c$Known T dwarf candidates identified by methane imaging  \citep{card15}. $^d$Known metal-poor T dwarf, HIP 73786B \citep{murr11}. $^e$T dwarf candidates that were not detected by the X-shooter $z$-band acquisition imaging. 
\end{list}
\end{table*}

\begin{table*}
 \centering
  \caption[]{Summary of the characteristics of the spectroscopic observations made with GTC/OSIRIS (first 3 rows) and VLT/X-shooter (last 15 rows). OSIRIS with R300R grism has wavelength coverage of 500--1000 nm. A 0.8 arcsec slit and R300R grism were used for the OSIRIS observations. Wavelength ranges for the VIS and NIR arms of X-shooter are 530--1020 and 990--2480 nm. A 1.2 arcsec slit are used for both VIS and NIR arms.  }
\label{tsdlob}
  \begin{tabular}{r l  c c c r r c c}
\hline
    Name~~~~~~~~~~~~  & SpT  & UT date & Seeing  & Airm & $T_{\rm int}$(VIS) &    $T_{\rm int}$(NIR) & Telluric (SpT) & Airm   \\
     & & &  (arcsec) & & (s)~~~~ & (s)~~~~ & &    \\   
\hline
ULAS J000844.34+012729.4 & T6 & 2015-08-23 & 0.70 & 1.13 & 2$\times$900 & --- & ---  & ---  \\
ULAS J002136.00+155227.3 & T4p & 2015-08-20 & 0.80 & 1.03 & 3$\times$900  & --- & --- & ---    \\
ULAS J012947.35+151143.1 & T4  & 2015-08-20 & 0.80 & 1.03 & 4$\times$900 & --- & --- & ---  \\
\hline
ULAS J000509.42+101407.7 & T3  & 2015-10-23 & 0.91 & 1.41 & 6$\times$290  & 7$\times$300  & HIP 9022 (B8/9 V) & 1.33  \\
ULAS J000844.34+012729.4 & T6.5$^a$  & 2015-10-28 & 1.03 & 1.14 & 4$\times$290  & 4$\times$300 & HIP 1191 (B8.5 V) & 1.07  \\
ULAS J002136.00+155227.3 & T4p  & 2015-10-29 & 1.09 & 1.34 & 8$\times$290  & 8$\times$300  & HD 216009 (A0 V) & 1.33 \\ 
ULAS J004030.41+091524.8 & T4.5 & 2015-10-29 & 1.08 & 1.21 & 8$\times$290  & 8$\times$300   & HIP 117927 (B9 V) & 1.23\\
ULAS J012252.58+043735.3 & T2.5  & 2015-10-30 & 2.08 & 1.23 & 12$\times$290  & 12$\times$300 & HIP 117927 (B9 V) & 1.21 \\
VIK J012834.94$-$280302.6  & T5   & 2016-01-26 & 1.13 & 1.54 & 4$\times$290  & 4$\times$300  & HIP 67796 (B2 II) & 1.68 \\
ULAS J014443.26+014741.0 & T5  & 2015-12-24 & 1.10 & 1.22 & 8$\times$290  & 8$\times$300  & HIP 3741 (B9 V) & 1.31 \\
ULAS J083346.88+001248.0 & T3  & 2016-01-29 & 0.88 & 1.24 & 8$\times$290  & 8$\times$300  & HD 77281 (A3 IV) & 1.22 \\
VIK J091641.18$-$010911.5 & T6  & 2016-01-28 & 1.19 & 1.49 & 6$\times$290  & 6$\times$300  & HIP 18788 (B5 III/IV) & 1.50  \\
ULAS J105131.80+025751.0 & T7  & 2016-02-20 & 0.66 & 1.15 & 6$\times$290  & 6$\times$300   & HR 3300 (A0 V) & 1.33 \\
ULAS J112059.40+121904.4 & T4.5  & 2016-03-02 & 0.87 & 1.42 & 8$\times$290  & 8$\times$300  & HIP 50480 (B9 V) & 1.42 \\
ULAS J124639.32+032314.2 & T5  & 2016-03-19 & 0.93 & 1.28 & 6$\times$290  & 6$\times$300  & HIP 43394 (B9 V) & 1.21 \\
VHS J132141.70$-$092446.2 & T4.5 & 2014-03-11 & 0.95 & 1.13 & 2$\times$1800 & 6$\times$600  & HIP 73881 (B2/3 V) & 1.10  \\
ULAS J135950.14+062553.9 & T3  &  2016-03-22 & 0.77 & 1.21 & 8$\times$290  & 8$\times$300   & HIP 87150 (B9/A0 III) & 1.19 \\
ULAS J150457.65+053800.8 & sdT5.5  & 2016-03-22 & 0.60 & 1.45 & 4$\times$290  & 4$\times$300  & HIP 79439 (B9 V) & 1.42 \\
\hline
\end{tabular}
\begin{list}{}{}
\item[] $^{a}$Candidate of unresolved BD binary of two T6+T8 dwarfs.  
\end{list}
\end{table*}

\section{Observations}
\label{sob}
\subsection{Candidate selection}
\label{scs}
T subdwarfs are extremely cool and have most of their flux at NIR wavelengths. We searched for T subdwarf candidates in the 10th data release of the UKIDSS Large Area Survey (LAS). Our searching criteria applied in LAS are $14.5 < J < 18.5$, $J-H < Y-J - 0.6$, $Y-J > 0.5$, and $J-K < 1.5(Y-J) - 0.7$ or $K$-band non-detection. These criteria are based on colours of latest L subdwarfs and known T dwarfs. We applied $|b| > 20$ deg to avoid the Galactic plane and requested non-detection in the Sloan Digital Sky Survey \citep[SDSS;][]{yor00} within 6 arcsec of LAS positions. We visually checked SDSS images to remove objects that were detected but not catalogued in SDSS due to light contamination from nearby bright stars. Then, remaining objects were cross-matched in the {\sl AllWISE} data base. We applied colour cuts $J-W2 > 1.75$ and $W1-W2 > 0.6$. We also removed objects that were discovered in UKIDSS with SDSS non-detection \citep{day13} and those having poor {\sl AllWISE} detection. With the same criteria, we selected some targets from the Visible and Infrared Survey Telescope for Astronomy's (VISTA) Hemisphere Survey \citep[VHS;][]{mcma12}  and Kilo-Degree Infrared Galaxy Survey \citep[VIKING;][]{suth12}, where overlapping with SDSS and visual inspection is possible. Table \ref{tdcand} shows our candidates that were followed up spectroscopically.

\subsection{GTC OSIRIS spectroscopy}
Three of our candidates were followed up with the Optical System for Imaging and low Resolution Integrated Spectroscopy \citep[OSIRIS;][]{cepa00} on the Gran Telescopio Canarias (GTC) in  2015 August. A 0.8 arcsec slit and a R300R grism were used on the OSIRIS observations. The OSIRIS R300R grism provides a resolving power of $\sim$300 and a wavelength coverage of 500--1000 nm. Our OSIRIS spectra were reduced with {\scriptsize IRAF} standard package\footnote{IRAF is distributed by the National Optical Observatory, which is operated by the Association of Universities for Research in Astronomy, Inc., under contract with the National Science Foundation.}. The spectral peak single-to-noise ratio (SNR) at $\sim$920 nm is 158 for UL0008, 21 for UL0021, and 33 for UL0129. Two white dwarf standard stars (Ross 640 and G 158-100) were used for the flux calibration. Telluric absorptions in these OSIRIS spectra were not corrected. 

\subsection{VLT X-shooter spectroscopy}
Our candidates are primarily followed up with X-shooter spectrograph \citep{ver11} on the Very Large Telescope (VLT). A total of 14 candidates (including UL0008 and UL0021 that were observed with the GTC) were observed between 2015 October and 2016 March. Another candidate was in the ESO online data base observed with X-shooter on 2014 March 11 (PI: M. Ruiz). X-shooter spectra were observed with a 1.2 arcsec slit in an ABBA nodding mode that provides a resolving power of 6700 in the visible (VIS) arm (530--1020 nm) and 4000 in the NIR arm (990--2480 nm). These spectra were reduced with ESO Reflex \citep{freu13}. Flux and wavelength calibrated 2D spectra were reduced to 1D spectra with the {\scriptsize IRAF} task {\scriptsize APSUM}. Telluric corrections were achieved using telluric standards that were observed right after or before our targets at a very close airmass. Note that five of our targets were too faint to be detected by the $z$-band acquisition imaging on X-shooter with 1--5 min exposure time, thus remain as candidates. 

\begin{figure*}
\begin{center}
  \includegraphics[width=\textwidth]{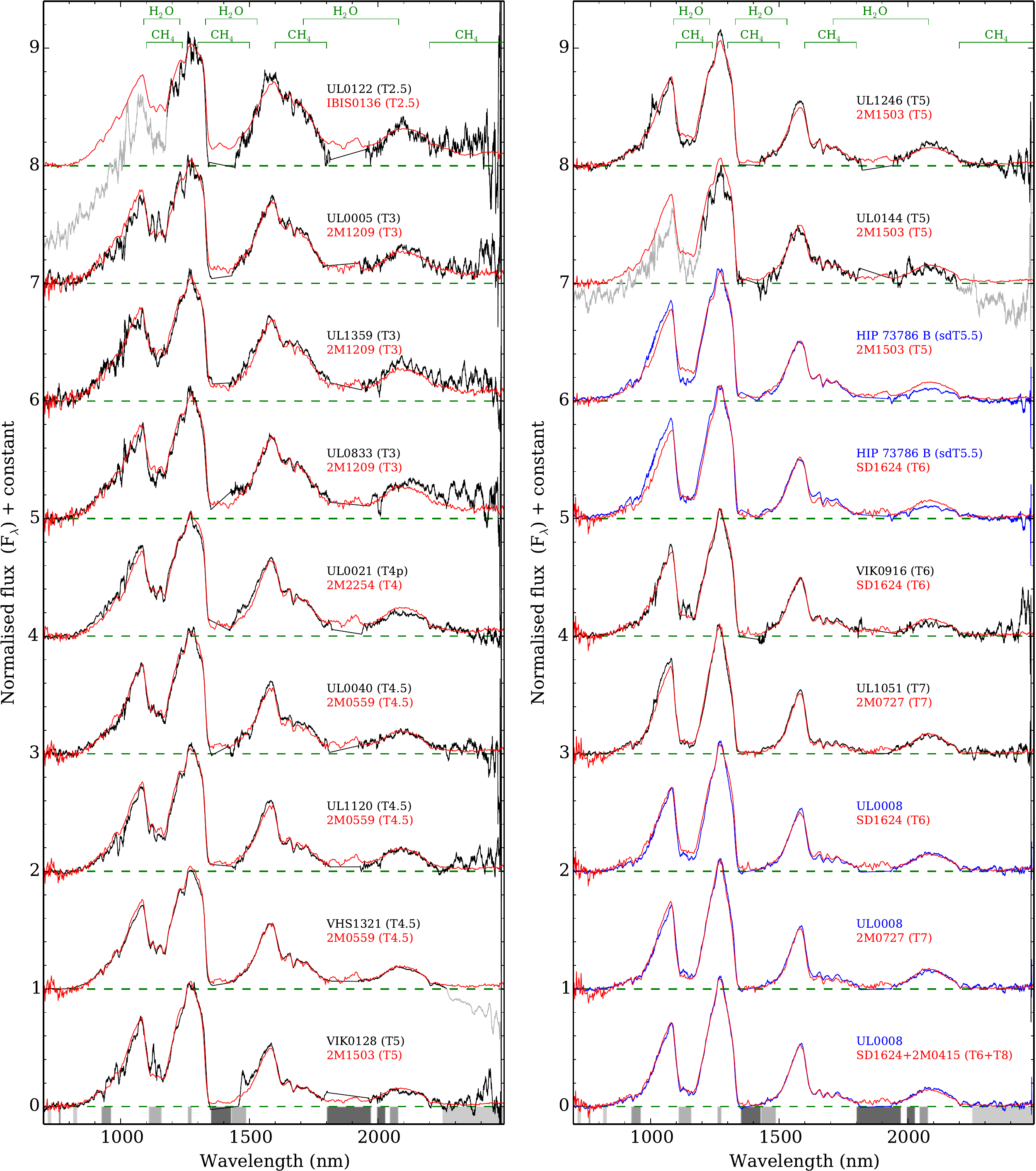}
\caption{X-shooter spectra of 15 T dwarfs observed with X-shooter on the VLT. Spectra are smoothed by 401 pixels in the VIS and 201 pixels in the NIR. Spectra of HIP 73786 B and UL0008 are highlighted in blue. Spectral wavelength of UL0008 was shifted by +300 km s$^{-1}$ for better fittings. Telluric absorption regions (corrected) are indicated with the shaded grey bands. The lighter and thicker shaded bands indicate regions with weaker and stronger telluric effects, respectively. Wavelength ranges in UL0122, VHS1321, and UL0144 that have flux calibration problem are plotted in grey. Standard T dwarf spectra used for comparison are IBIS0136 \citep{bur08}, 2MASS J12095613-1004008 (2M1209), 2MASS J15031961+2525196 (2M1503), 2MASSI J2254188+312349, \citep[2M2254;][]{bur04b}, 2MASS J05591914-1404488 \citep[2M0559;][]{bur06}, 2MASSI J0727182+171001 (2M0727), SD1624 \citep{bur06b}, 2M0415 \citep{bur04b}. Spectra are normalized at 1280 nm. }
\label{sdtvlt}
\end{center}
\end{figure*}

\begin{figure}
\begin{center}
  \includegraphics[width=\columnwidth]{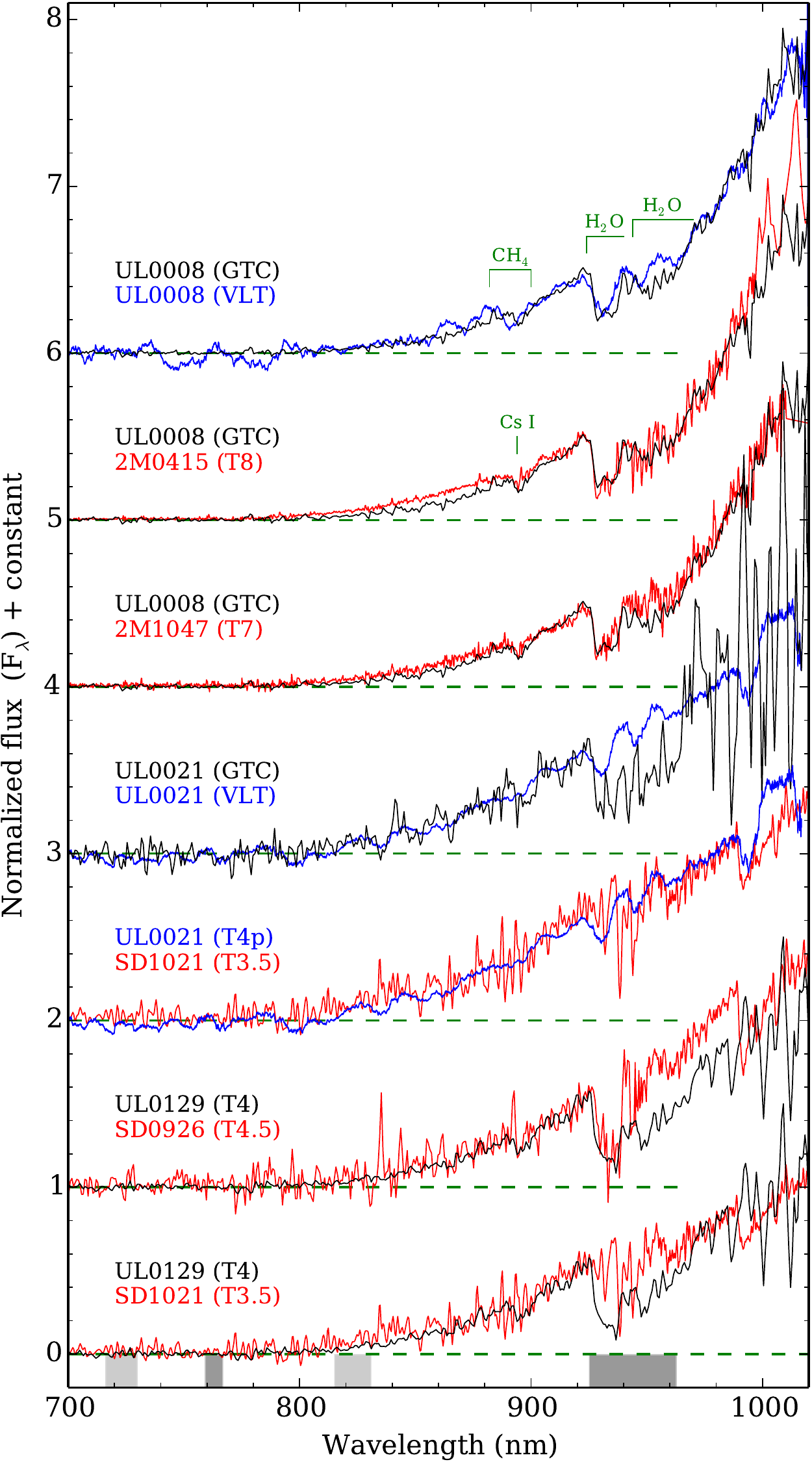}
\caption{Optical spectra of three new T dwarfs observed with X-shooter (blue) and OSIRIS (black). Spectra of 2M0415 and 2M1047 are from  \citet{bur03a}, and SDSSp J092615.38+584720.9 (SD0926) and SDSSp J102109.69$-$030420.1 (SD1021) are from \citet{geba02}.}
\label{sdtgtc}
\end{center}
\end{figure}

\begin{figure}
\begin{center}
  \includegraphics[width=\columnwidth]{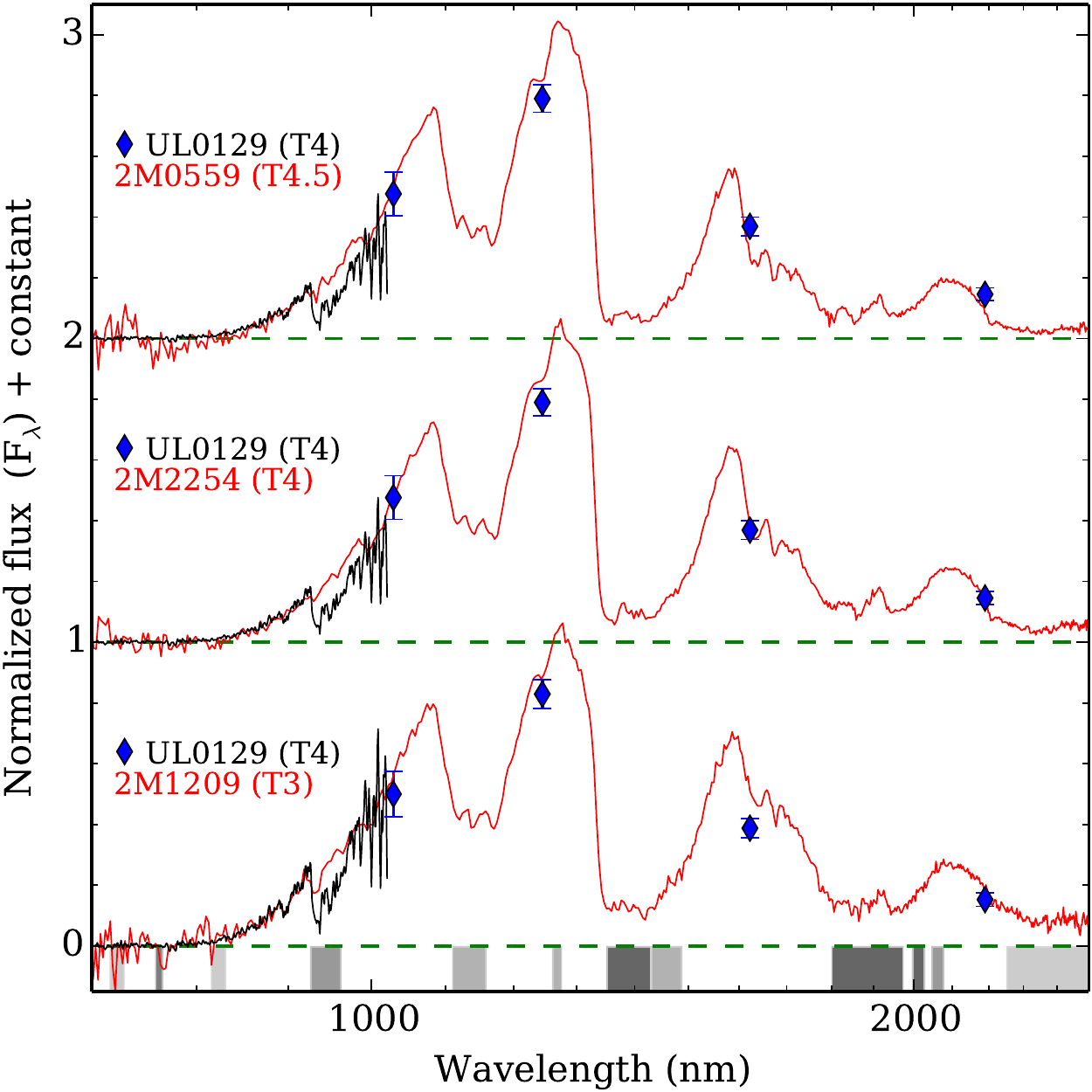}
\caption{The optical to NIR SED of UL0219. The blue diamonds are converted from UKIDSS photometry of UL0219. Spectra of 2M1209 and 2M2254 are from \citet{bur04b}, and 2M0559 is from \citet{bur06}. }
\label{j0129}
\end{center}
\end{figure}

\section{Classification} 
\label{ssc}
T dwarfs are primarily classified in the NIR \citep{burg02}. \citet{bur03a} presented an optical spectral sequence of T dwarf standards. T dwarfs have very red optical spectra and barely have flux at wavelength shorter than 800 nm. We classified our OSIRIS and X-shooter spectra by compare their optical and NIR spectra to those of standard T dwarfs. 

Fig. \ref{sdtvlt} shows the NIR spectra of 15 T dwarfs observed with X-shooter compared to standard T dwarf spectra normalized to their $J$- and $H$-band flux. Spectra displayed in Fig. \ref{sdtvlt} were smoothed by 401 (in VIS) and 201 (in NIR) pixels to increase their SNR by $\sim$20 and 14 times. Fig. \ref{sdtgtc} shows the optical spectrum of three T dwarfs observed with OSIRIS. 

\subsection{Two T dwarfs previously identified by methane imaging}
ULAS J112059.40+121904.4 (UL1120)  and ULAS J124639.32+032314.2 (UL1246) were previously identified as T dwarfs by  methane imaging, and were classified as T5.5 and T5 types, respectively \citep{card15}. We updated the spectral-type for UL1120 to T4.5 as its NIR spectra fits well to the T4.5 dwarf 2MASS J05591914$-$1404488 \citep[2M0559;][]{burg00}. UL1246 fits well to 2MASS J15031961+2525196 \citep[2M1503;][]{bur03b}, which has an NIR spectral-type of T5 \citep{bur06}, consistent with its spectral-type based on methane imaging.  

\subsection{Two T dwarfs previously identified by photometry}
ULAS J000844.34+012729.4 (UL0008) and ULAS J105131.80+025751.0 (UL1051) were previously identified as T6.5 dwarfs by multiband photometry \citep{skrz16}. We re-classified UL1051 as T7 as its NIR spectrum fits well to that of the T7-type 2MASSI J0727182+171001 \citep[2M0727;][]{burg02} in Fig. \ref{sdtvlt}. 
 
 Fig. \ref{sdtvlt} shows that UL0008 has an NIR spectral-type between the T6 SDSSp J162414.37+002915.6 \citep[SD1624;][]{stra99,bur06} and the T7-type 2MASSI J0727182+171001 \citep[2M0727;][]{burg02}. However, UL0008 does not fit very well to either the T6 or the T7 standard. UL0008 has a narrower $J$-band spectrum and deeper CH$_4$ absorption bands around 1150 and 1680 nm than T6. Meanwhile, UL0008 has more flux at the 1680 nm CH$_4$ absorption band but less flux around 1050 nm than T7. 

\subsection{UL0008 as a T6+T8 binary}
We observed the optical spectrum of UL0008 with OSIRIS before the X-shooter run. Fig. \ref{sdtgtc} shows that UL0008 has a similar red optical spectrum as the T7-type 2MASSI J1047538+212423 (2M1047) and the T8-type 2MASSI J0415195-093506  \citep[2M0415;][]{bur03a}, but has less flux around 800--900 nm wavelength. Note that the flux calibration of OSIRIS optical spectrum of UL0008 is consistent with its X-shooter VIS spectrum. Considering that UL0008 also cannot be fitted very well with single NIR spectra. We fitted the NIR of UL0008 with combined spectra of  SD1624 (T6) and 2M0415 \citep{bur04b}. Fig. \ref{sdtvlt} shows that UL0008 can be better fitted by a combined T6 + T8 spectrum than a single spectrum, thus is possibly an unresolved BD binary. 
 
\subsection{Updated spectral-type of a known metal-poor T dwarf}
ULAS J150457.65+053800.8 (HIP 73786B) was identified as a metal-poor T6 dwarf companion to HIP 73786, which is a K8V star with known metallicity of [Fe/H] = $-0.3\pm0.1$ \citep{murr11}.  HIP 73786B was one of our candidates, and we obtained its optical-NIR spectrum that has a higher quality (Fig. \ref{sdtvlt}). The methane absorption band near 1650 nm in HIP 73786B is shallower than the T6 standard SD1624. The $J$ and $H$ band spectrum of HIP 73786B fits well to the T5 standard 2M1503, but have relatively deeper absorption band near 1150 nm. HIP 73786B clearly has suppressed $K$-band flux and metallicity (same as HIP 73786 A) in the range of sdM and sdL subclasses (fig. 21; \citetalias{zha17a}). Therefore, we updated its spectral-type to sdT5.5. 

\subsection{A mildly metal-poor T dwarf}

ULAS J002136.00+155227.3 (UL0021) fits well to the T4 standard 2M2254 (Fig \ref{sdtvlt}). However, UL0021 has slightly less $K$-band flux than 2M2254 (Fig. \ref{sdtvlt}). The stronger suppression of $K$-band spectrum of UL0021 indicating that it might be mildly metal-poor. The OSIRIS optical spectrum of UL0021 is shown in Fig. \ref{sdtgtc}.  

\subsection{Ten new T dwarfs}
ULAS J012252.58+043735.3 (UL0122) was classified T2.5 (Fig. \ref{sdtvlt}) as its 1200--2300 nm wavelength spectrum fits well to the T2.5 standard, IBIS J013656.57+093347.3 \citep[IBIS0136;][]{arti06}. Note that the flux calibration of UL0122 around 700--1200 nm is not correct, possibly due to its low SNR. 

ULAS J000509.42+101407.7 (UL0005), ULAS J083346.88+001248.0 (UL0833), and ULAS J135950.14+062553.9 (UL1359) were classified T3 as they fit well to 2MASS J12095613$-$1004008 \citep[2M1209;][]{bur04b}, which has a T3 NIR type and a T3.5 optical type \citep{kirk08}. 

ULAS J004030.41+091524.8 (UL0040) and VHS J132141.70$-$092446.2  (VHS1321) both fit well to the T4.5 (NIR) type 2MASS J05591914$-$1404488 \citep[2M0559;][]{burg00} therefore are classified as T4.5. Note that the flux calibration of VHS1321 around 2250--2500 nm is not correct, possibly due to its low SNR.

VIK J012834.94$-$280302.6 (VIK0128) and ULAS J014443.26+014741.0 (UL0144) fit well to the T5 (NIR) type 2MASS J15031961+2525196 \citep[2M1503;][]{bur03b} and are classified as T5.  Note that the flux calibration of UL0144 around 700--1200 and 2180--2500 nm is not correct, possibly due to its low SNR.

VIK J091641.18$-$010911.5 (VIK0916) fits well to the T6 standard SD1624 and is classed as T6. Note it has slightly less $K$-band flux, but could be caused by noise. 

ULAS J012947.35+151143.1 (UL0129) was observed with OSIRIS (Fig. \ref{sdtgtc}). Its optical spectrum is similar to that of the T3.5-type SDSSp J102109.69$-$030420.1 \citep[SD1021;][]{legg00} and the T4.5-type SDSSp J092615.38+584720.9 \citep[SD0926;][]{geba02}.  Fig. \ref{j0129} shows the optical to NIR spectral energy distribution (SED) of UL0129 compared to the T3-type 2M1209, the T4-type 2MASSI J2254188+312349, \citep[2M2254;][]{burg02}, and the T4.5-type 2MASS J05591914$-$1404488 \citep[2M0559;][]{burg00}. The optical spectrum and NIR SED of UL0129 are consistent with a spectral classification of T4 type. 

\section{The physics of brown dwarfs}
\label{sebd}
\begin{figure}
\begin{center}
  \includegraphics[width=\columnwidth]{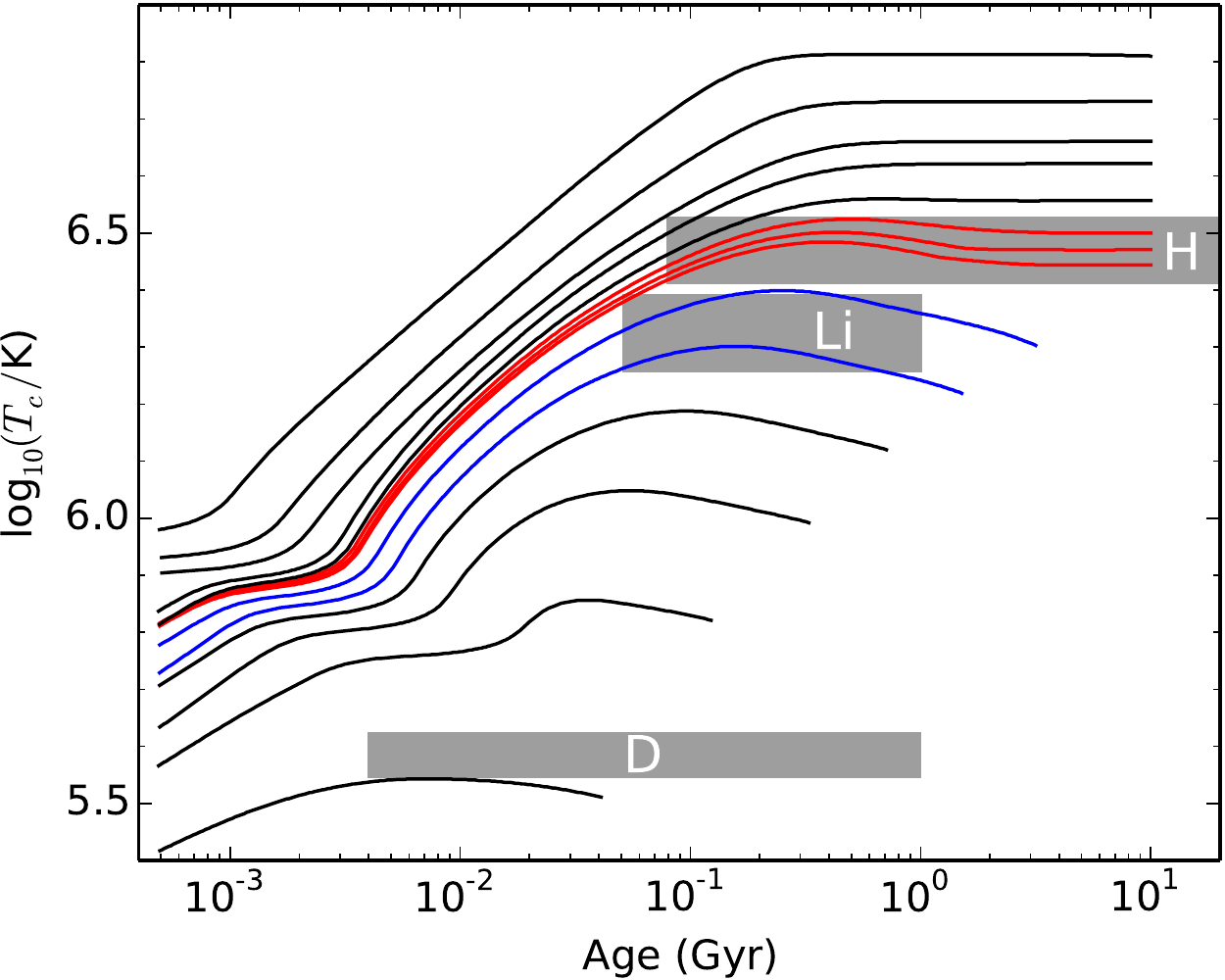}
\caption{Evolutionary tracks of core temperature ($T_{\rm c}$) of objects with mass of 0.01, 0.02, 0.03, 0.04, 0.05 (blue), 0.06 (blue), 0.07 (red), 0.072 (red), 0.075 (red), 0.08, 0.09, 0.1, 0.13, and 0.2 M$_{\sun}$ \citep{bara15}. The grey shaded areas indicate the $T_c$ and age ranges that the unsteady hydrogen fusion take place (top), and lithium (middle) and deuterium (bottom) nuclear fusions were incomplete. The age range of unsteady hydrogen fusion is based on fig. 6 of \citet{burr11}. The age ranges of incomplete fusions of deuterium and lithium are based on \citet[figs 6 and 7]{burr01} and Fig. \ref{fli}.  
}
\label{ftc}
\end{center}
\end{figure}

The theory of BDs was extensively discussed in the literature \citep[e.g.][]{burr93b,chab97,burr01}. 
BDs are fully convective objects and most of them rely on their initial thermal energy. However, they do have nuclear reactions to fuse hydrogen, deuterium, and lithium in their cores. Fig. \ref{ftc} shows the evolution of core temperature ($T_{\rm c}$) of objects with mass between 0.01 and 0.2 M$_{\sun}$. BDs in the STZ (T-BDs) could reach temporary states of $T_{\rm c}$ and pressure to fuse hydrogen at low levels in their cores. Massive BD above $\sim0.05M_{\sun}$ could fuse lithium at young age. BDs with mass above $\sim$0.01$M_{\sun}$ could fuse deuterium at young age. 

The different evolutions of T-BD and D-BDs caused by unsteady hydrogen fusion are well revealed by isochrones of mass--$T_{\rm eff}$ and mass--luminosity \citep[e.g.][]{burr93}. We compared $T_{\rm eff}$ isochrones of very low-mass stars (VLMS) and BDs predicted by evolutionary models \citep{burr97,bara03,bara15} to observational measurements \citep{dupu17,lazo18} in our previous work (fig. 20; \citetalias{zha18b}). It seems that the $T_{\rm eff}$ of VLMS just above the STZ, around 0.085--0.11 M$_{\sun}$ are overestimated by \citet{burr97} according to observational data. Isochrones predicted by \citet{bara03,bara15} are relatively well consistent with observations. 

\subsection{Hydrogen burning transition zone}
\label{shbtz}
\begin{figure*}
\begin{center}
  \includegraphics[width=\textwidth]{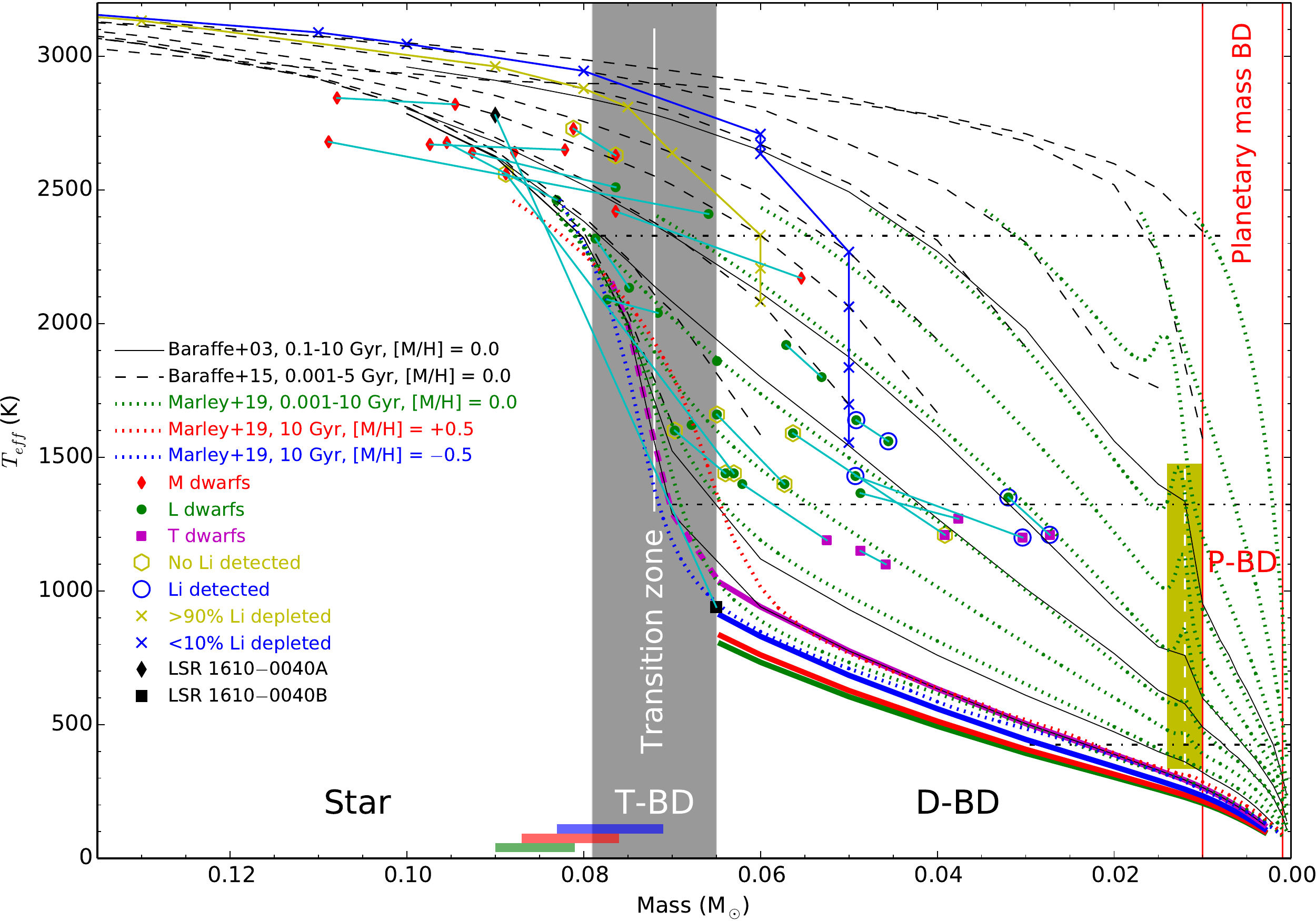}
\caption{Isochrones of $T_{\rm eff}$ of VLMS and substellar objects. The black solid lines are solar metallicity isochrones at 0.1, 0.5, 1, 5, and 10 Gyr \citep{bara03}. The black dashed lines are solar metallicity isochrones at 0.001, 0.01, 0.05, 0.1, 0.2, 0.3, 0.5, 1, and 5 Gyr  \citep{bara15}. The green dotted lines are solar metallicity isochrones at 0.001, 0.01, 0.04, 0.1, 0.2, 0.4, 1, 2, 4, and 10 Gyr (Marley et al. 2019, in prep.). The 10 Gyr isochrones with [M/H] = +0.5 and [M/H] = $-0.5$ (i.e. [Fe/H] $\approx -0.7$) from Marley et al. (2019) are overplotted and intersect with the 10 Gyr isochrone with [M/H] = 0.0 at 0.079 M$_{\sun}$, which defines the stellar-substellar boundary. A grey shaded band indicates the STZ (at $Z_{\sun}$) that hosts T-BDs, and have stars to its left and D-BDs to its right. A white line in the middle of the transition zone indicates the HBMM (0.072 M$_{\sun}$) predicted by \citet{chab97}. The red diamonds, the green circles, and the magenta squares are M7--T5 dwarfs that are in binary systems (joined with the cyan lines) thus have measurements of dynamical masses \citep{dupu17,lazo18}. Masses of an unresolved subdwarf binary, LSR 1610$-$0040AB (the black diamond and square), is from \citep{kore16}. Objects with and without lithium detection in their spectra are indicated with blue circles and yellow hexagons, respectively. The yellow and blue crosses indicate  $>90$ per cent and $<10$ per cent lithium depletion predicated by evolutionary models \citep{bara15}. The three dot-dashed lines indicate the average $T_{\rm eff}$ of  L0, T0, and Y0 dwarfs \citep{dupu13,dupu17}. The two vertical red lines indicate 1M$_{\rm Jup}$ and 0.01 M$_{\sun}$ (deuterium burning minimum mass). A yellow shaded band on the right indicates the deuterium burning transition zone (0.01--0.014 M$_{\sun}$), centred at 0.012 M$_{\sun}$ (the white dashed line). The 10 Gyr isochrones for T-BDs and D-BDs are highlighted with the magenta dashed and solid thick lines, respectively. The blue, red, and green solid thick lines indicate the 10 Gyr isochrones at 0.1, 0.01, and 0.001 $Z_{\sun}$, respectively. The blue, red, and green bars at the bottom indicate mass ranges of STZ at 0.1, 0.01, and 0.001 $Z_{\sun}$.
}
\label{fiso}
\end{center}
\end{figure*}

Fig. \ref{fiso} shows the solar metallicity isochrones of VLMS, T-BDs, D-BDs, and planetary mass objects from 0.001 to 10 Gyr. M7--T5 dwarfs in binary systems with measured dynamical masses are overplotted for comparison \citep{dupu17,lazo18}. The hydrogen burning transition zone is indicated with a grey shaded band at around 0.065--0.079 M$_{\sun}$, where isochrones are stretched from 1 to 10 Gyr. The 10 Gyr isochrone with [M/H] = 0.0 intersects with the 10 Gyr isochrones with [M/H] = +0.5 and [M/H] = $-0.5$ (Marley et al. 2019, in prop.) at 0.079 M$_{\sun}$, which defines the stellar-substellar boundary at [M/H] = 0.0. Meanwhile, the gravity maximum and radius minimum of dwarfs with [M/H]= 0.0 and 10 Gyr age are around 0.065 M$_{\sun}$ according to evolutionary models (Marley et al. 2019), which defines the boundary between T-BDs and D-BDs. The HBMM predicted by \citet{chab97} around 0.072 M$_{\sun}$ is actually the middle of the hydrogen burning transition zone.  

Hydrogen is the most abundant element in BDs, and unsteady hydrogen fusion in T-BDs could last for $\gg$ 10 Gyr \citep[e.g. fig. 6;][]{burr11}. Therefore, the hydrogen burning transition zone is also the STZ that separates stars, T-BDs, and D-BDs. D-BDs are below the STZ and do not have hydrogen fusion, thus cool much faster than T-BDs. Spectral-type and $T_{\rm eff}$ range of T-BDs in the STZ are stretched between stars and D-BDs with the dissipation of initial thermal energy and steeply reduced efficiency of nuclear fusion towards lower mass and $T_{\rm c}$. 

With about 10 Gyr of cooling, halo T-BDs has more distinct properties from VLMS and D-BDs. 
2MASS J05325346+8246465 \citep[2M0532;][]{burg03} is the first L subdwarf identified. It has $T_{\rm eff}$ = 1600 K and [Fe/H] = $-$1.6, and is classified as esdL7 type \citepalias{zha17b}. About half of its luminosity is contributed from hydrogen fusion \citep{burg08c}, thus 2M0532 is a T-BDs in the STZ. The lack of L3--T5 subdwarfs is revealed by simulated evolutions of VLMS and BDs with halo distributed birth rate \citep[fig. 10;][]{burg04}. There are only 10 halo T-BDs known in the literature (\citetalias{zha18b,zha19}). Four BD candidates were detected in the Messier 4 (M4) globular cluster with [Fe/H] $\approx -1.1$ at a distance of $\sim$2 kpc \citep{dieb16}. These three brighter ones are possibly T-BDs, and the faintest one is probably a D-BD according to fig. 9 of \citetalias{zha17b}.

The LSR 1610$-$0040AB \citep{lepi03} is an unresolved binary \citep{dahn08} that composed of a VLMS and a D-BD seaprated by the STZ. LSR 1610$-$0040AB barely have VO absorption band at 800 nm, which is a criterion for esd sublcass with [Fe/H] $\la -1.0$ \citepalias{zha17a}.  \citet{kore16} shows that LSR 1610$-$0040AB, with [Fe/H] = $-$1.0, would have masses of 0.09 and 0.065 M$_{\sun}$, and corresponding $T_{\rm eff}$ of 2780 K and 940 K at 10 Gyr according to \citetalias{zha17b} (fig. 9) and \citetalias{zha18a} (fig.5), respectively. LSR 1610$-$0040AB would be a esdM8+esdT6 binary by the spectral-type -- $T_{\rm eff}$ correlation \citepalias[fig. 4;][]{zha18a}. LSR 1610$-$0040A have subsolar metallicity thus has higher $T_{\rm eff}$ than the solar-metallicity 10 Gyr isochrone at 0.09 M$_{\sun}$ in Fig. \ref{fiso}. For the same reason \citep[fig. 5;][]{burr01}, LSR 1610$-$0040B has lower $T_{\rm eff}$ than the solar-metallicity 10 Gyr isochrone at 0.065 M$_{\sun}$, but consistent with the shifted isochrone for 0.1 $Z_{\sun}$.

D-BDs, T-BDs, and VLMS in the field are more difficult to distinguish without knowing their ages. GD 165B (L4) is the first L dwarf \citep{beck88,kirk93}, and is probably a T-BD \citep[fig. 5;][]{kirk99}. The first M-type BD, Teide 1 \citep[M8;][]{rebo95} is a D-BD, as it is not massive enough to fuse lithium \citep{rebo96}.

\begin{figure}
\begin{center}
  \includegraphics[width=\columnwidth]{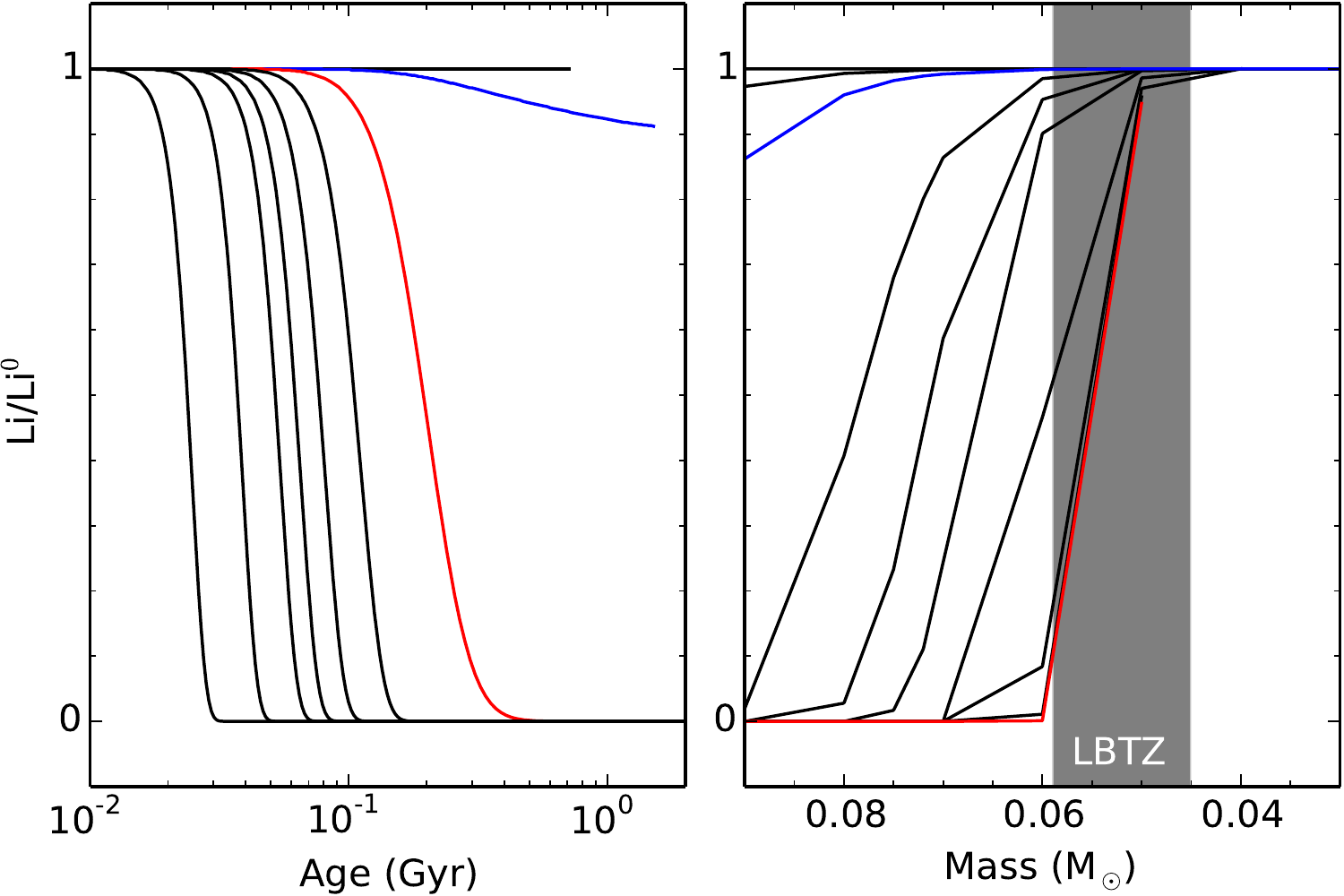}
\caption{The evolutionary tracks (left) and isochrones (right) of the ratio of surface lithium abundance to initial abundance (Li/Li$^0$) due to nuclear fusion in very low-mass objects \citep{bara15}. The masses of these evolutionary tracks are 0.04, 0.05 (blue), 0.06 (red), 0.07, 0.08, 0.09, 0.13, and 0.2 M$_{\sun}$. The age of these isochrones are 0.02, 0.04, 0.05 (blue), 0.08, 0.1, 0.12, 0.2, 0.3, 0.4, 0.5 (red) Gyr. The grey band indicates the mass range of lithium burning transition zone (LBTZ) with incomplete lithium fusion. 
}
\label{fli}
\end{center}
\end{figure}

\begin{figure}
\begin{center}
  \includegraphics[width=\columnwidth]{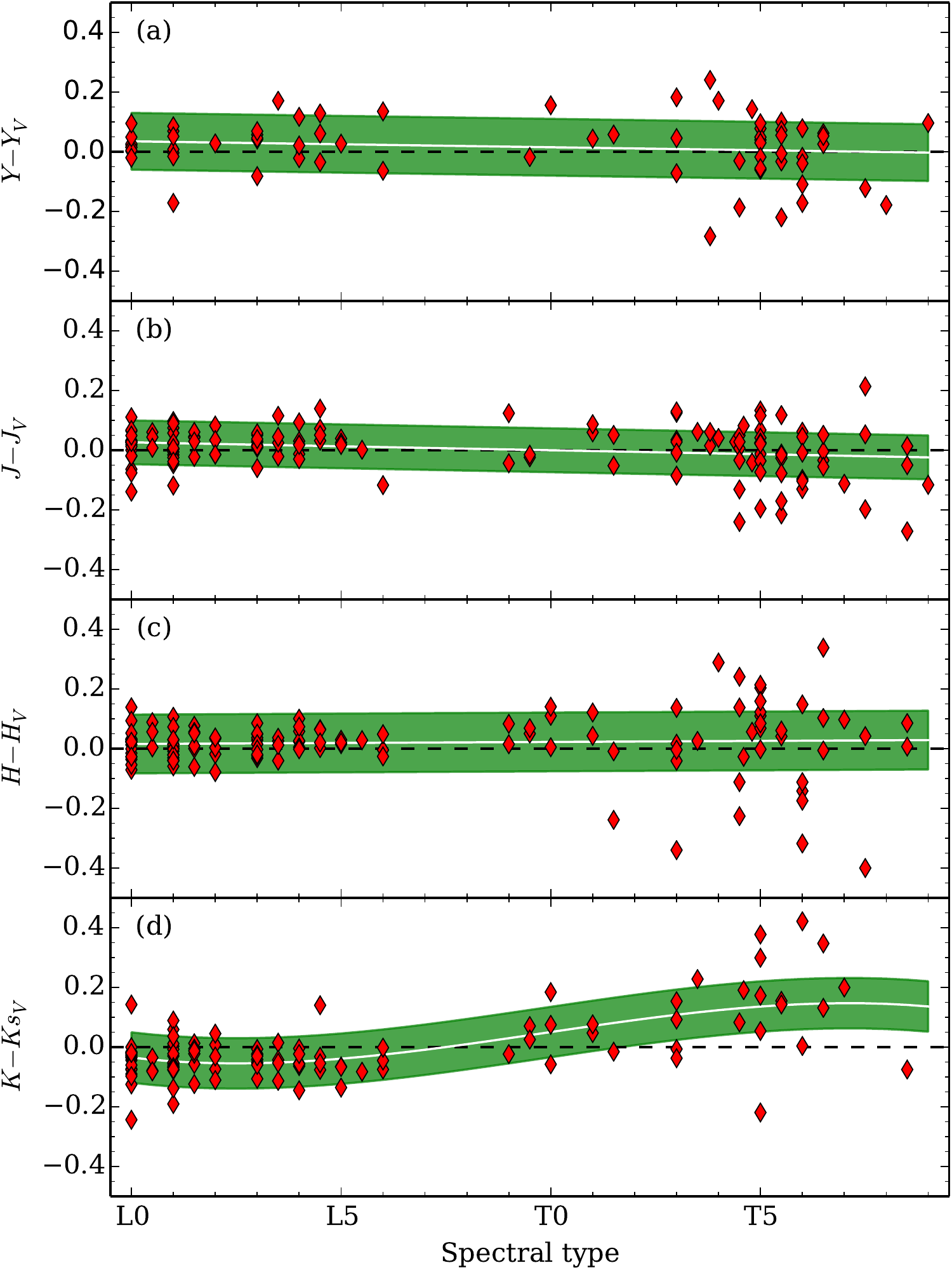}
\caption{Polynomial fits (the white lines) to UKIDSS--VISTA magnitude differences as a function of spectral-type for L0--T9 dwarfs observed in LAS and VHS/VIKING (the red diamonds). The green shaded area indicates the rms of these fits. 
The black dashed lines are zero-points. 
}
\label{fyjhk}
\end{center}
\end{figure}

\subsection{Lithium burning transition zone}
Fig. \ref{fli} shows the evolutionary tracks and isochrones of the ratio of surface lithium abundance to initial abundance (Li/Li$^0$) in the atmospheres of BDs and stars. The  Li/Li$^0$ of BDs with 0.05 M$_{\sun}$ slowly declined from 1 to 0.92 at the age between 0.1 and 1 Gyr. While, the Li/Li$^0$ of BDs with 0.06 M$_{\sun}$ quickly declined from 0.95 to 0.08 at the age between 0.1 and 0.3 Gyr, and disappeared before 0.6 Gyr. A lithium burning transition zone (LBTZ) of a mass range from $\sim$0.045 M$_{\sun}$ to slightly below 0.06 M$_{\sun}$ exists among D-BD population. The lithium nuclear reaction is incomplete in D-BDs of this mass range at the age between 0.05 and 1 Gyr, before their core temperatures are too low to fuse lithium or their atmospheres are cool enough ($T_{\rm eff} \la 1525$ K) to condense lithium into LiCl, LiOH, and LiH \citep{lodd99}. Lithium in D-BDs that below the LBTZ ($\la$ 0.045 M$_{\sun}$) starts to condense at slightly earlier ages (cool faster), and likely disappear before an age of $\sim$2 Gyr (Fig. \ref{fiso}). Note that the LBTZ predicted by \citet[fig. 7]{burr01} is around 0.05--0.065 M$_{\sun}$, which is about 0.005 M$_{\sun}$ higher than that of \citet{bara15}. 

The yellow and blue lines/crosses in Fig. \ref{fiso} mark where $>$ 90\% and $<$ 10\% lithium are depleted, respectively. L and T dwarfs without and with lithium detections in their spectra are highlighted with the yellow hexagons and the blue circles, respectively. Luhman 16B \citep{luhm13} is the coolest and lowest mass object with direct lithium detection \citep{lodi15} at an age of $\sim$0.5 Gyr.

The LBTZ is well below the STZ. Therefore, the lithium depletion test \citep{maga93} can be used to identify massive D-BDs, but only for late-M \citep[e.g. PPL 15;][]{basr96} and L \citep[e.g. Kelu-1;][]{ruiz97} dwarfs younger than $\sim$2 Gyr and early T-type dwarfs (e.g. Luhman 16B) younger than $\sim$0.8 Gyr. Monatomic lithium is not expected in the atmospheres of metal-poor T-BDs and D-BDs, which either are massive enough to fuse or cool enough to condense lithium  \citep[section 3.5;][]{zha18a}. 

\subsection{Deuterium burning transition zone}
A deuterium burning transition zone exists around 0.01--0.014 M$_{\sun}$ according to those bumps on isochrones at 0.1--1 Gyr (Fig. \ref{fiso}) calculated by \citet{bara03}. The cooling of BDs in this mass range is slowed down due to extra energy supply when their $T_{\rm c}$ is high enough to fuse deuterium at the age of $\sim$0.004--1 Gyr (Fig. \ref{ftc}). The deuterium burning transition zone is also revealed by the evolutionary tracks of deuterium abundance due to nuclear fusion in substellar objects around 11--15 $M_{\rm Jup}$ \citep[fig. 6;][]{burr01}, which is slightly higher than that predicted by \citet{bara03}. 

D-BDs with masses above the deuterium burning transition zone reach the minimum $T_{\rm c}$ to fuse deuterium at earlier ages, thus burn all deuterium at earlier ages (e.g. $\la$0.02 Gyr for 0.02 M$_{\sun}$). A BD with 0.02 M$_{\sun}$ has $T_{\rm eff}$ above 2500 K at 0.02 Gyr. The fraction of its energy contributed from deuterium fusion is tiny, thus not revealed in Fig. \ref{fiso}. The cooling of BD with mass of 0.015 M$_{\sun}$ seems to slow down at 0.05 Gyr, which is caused by deuterium fusion. 

Deuterium fusion requires the lowest temperature/pressure among nuclear fusions in celestial objects. The deuterium burning transition zone could be used to define the upper mass boundary of exoplanets. Note that the $\sim$0.012 M$_{\sun}$ \citep{bara03} or the 13$M_{\rm Jup}$ \citep{burr01} used in the literature is middle of the deuterium burning transition zone. The actual deuterium burning minimum mass is at the bottom of the transition zone, around 0.01 M$_{\sun}$ ($\sim$10.5 M$_{\rm Jup}$).

P-BDs as a subgroup of D-BDs, are useful in the analogue study of gaseous exoplanets. Furthermore, the deuterium burning transition zone is a theoretical mass range, and may not possible to reveal by direct observation. The deuterium burning transition zone have very weak impact on the properties of BDs. Moreover, D-BDs and gaseous exoplanet are very similar populations, as they follow the same mass--radius and mass--density correlations \citep{hatz15}.

\subsection{Metal-poor degenerate brown dwarfs}
\label{sdbd}
Most of the D-BDs would have crossed the $T_{\rm eff}$ range of the STZ and become T5+ types ($\la$ 1000 K) in $\sim$5 Gyr (Fig. \ref{fiso}) according to evolutional models \citep{burr01,bara03}. Consequently, the vast majority of T0--4 dwarfs are D-BDs younger than $\sim$3  Gyr.
VLMS with subsolar metallicity are hotter than those with solar metallicity, and D-BDs with subsolar metallicity are cooler than those with solar metallicity according to isochrones of VLMS and BDs \citep[fig. 5;][]{burr01}.  To indicate the 10 Gyr isochrones of D-BDs at metallicities of 0.1, 0.01, and 0.001 $Z_{\sun}$, we shifted  the 10 Gyr solar metallicity isochrone down by 12\%, 19\%, and 22\% in Fig. \ref{fiso} (the blue, red, green thick solid lines). The shifts are calculated from these 10 Gyr isochrones of D-BDs with metallicities of 1.0, 0.1, 0.01, and 0.001 $Z_{\sun}$ in figure 5 of \citet{burr01}. The mass range of the STZ at subsolar metallicity is higher than that at $Z_{\sun}$. The STZ is between 0.071 and 0.083 M$_{\sun}$ at 0.1 $Z_{\sun}$ ([Fe/H] = $-1.3$), 0.076 and 0.087 M$_{\sun}$ at 0.01 $Z_{\sun}$ ([Fe/H] = $-2.3$), and 0.081--0.09 M$_{\sun}$ at 0.001 $Z_{\sun}$ ([Fe/H] = $-3.3$) according to fig. 5 in \citetalias{zha18a}.  Fig. \ref{fiso} shows that all metal-poor D-BDs would have spectral-types of sdT5 or later, after $\ga$ 8 Gyr of cooling. The ratio between T5--9 and T0--4 subdwarfs would be very high.  Because T0--4 subdwarfs are T-BDs in a narrow mass range on the initial mass function. Their spectral-type sampling is stretched by the STZ and further stretched by the rapid evolution of atmospheres at L/T transition (section 4.4; \citetalias{zha18b}).

\begin{table*}
 \centering
  \caption[]{Photometry of known metal-poor T dwarfs in the literature that are likely in the same metallicity range as sdM and sdL subclasses \citepalias{zha17a}. }
\label{tkmptd}
  \begin{tabular}{l l  c c c c c c c}
\hline
    Name & SpT$^a$ & $Y$ & $J$ & $H$ & $K$ & $W1$ & $W2$ & Ref.$^c$ \\
\hline
WISE J001354.39+063448.2 & T8p & --- & 19.75$\pm$0.12 & 20.02$\pm$0.12 &--- & $\ge$18.23  & 15.23$\pm$0.09 & 15 \\
ULAS J012855.07+063357.0 & T6 & 19.66$\pm$0.14 & 18.93$\pm$0.12 & --- & --- &--- &--- & 4 \\
WISE J013525.64+171503.4 & T6 &--- & 17.03$\pm$0.02 & --- &--- & 16.89$\pm$0.10 & 14.64$\pm$0.06 & 9 \\
WISE J023318.05+303030.5$^b$ & T6 &--- & 16.86$\pm$0.34 &--- &--- & 16.40$\pm$0.07 & 14.34$\pm$0.05 & 9  \\
ULAS J025545.28+061655.8 & T6 & 19.15$\pm$0.07 & 17.99$\pm$0.05 & 18.67$\pm$0.18 & --- & 17.73$\pm$0.19 & 15.27$\pm$0.09 & 4, 11 \\
WISE J044853.28$-$193548.6 & T5p & --- & 16.61$\pm$0.01 & --- & 17.58$\pm$0.14 & 16.29$\pm$0.06 & 14.24$\pm$0.04 & 6 \\
WISE J052536.35+673952.6 & T6p &--- & 17.49$\pm$0.04 & 18.87$\pm$0.05 &--- & 16.79$\pm$0.09 & 14.79$\pm$0.06 & 6 \\
WISE J052844.51$-$330823.9 & T7p & --- & 16.90$\pm$0.07 & 16.97$\pm$0.14 & 17.44$\pm$0.20 & 17.23$\pm$0.11 & 14.52$\pm$0.05 & 6 \\
WISE J061213.85$-$303612.5AB & T6 &--- & 16.65$\pm$0.03 & 16.96$\pm$0.03 &--- & 16.40$\pm$0.06 & 14.04$\pm$0.04 & 6 \\
WISE J061407.49+391235.9 & T6 &--- & 16.9$\pm$0.363 &--- &--- & 16.24$\pm$0.08 & 13.63$\pm$0.03 & 6 \\
2MASS J07290002$-$3954043 & T8p & 16.52$\pm$0.08 & 15.64$\pm$0.08 & 16.05$\pm$0.18 &--- &--- &--- & 8 \\
WISE J083337.83+005214.2 & T9p & --- & 20.28$\pm$0.14 & 20.63$\pm$0.10 &--- & $\ge$18.34  & 14.97$\pm$0.08 & 15 \\
WISE J083641.10$-$185947.0 & T8p & --- & 19.10$\pm$0.12 & --- & --- & 17.62$\pm$0.18 & 15.15$\pm$0.07 & 6 \\
ULAS J090116.23$-$030635.0 & T7.5 & 18.90$\pm$0.05 & 17.89$\pm$0.04 & 18.49$\pm$0.13 & --- & 17.19$\pm$0.13 & 14.56$\pm$0.05 & 7 \\
ULAS J092744.20+341308.7 & T5.5 & 19.66$\pm$0.14 & 18.77$\pm$0.11 & --- & --- & 16.04$\pm$0.06 & 16.06$\pm$0.17 & 4 \\
2MASS J09373487+2931409 & T6p &--- & 14.30$\pm$0.01 &14.67$\pm$0.03 & 15.39$\pm$0.06 & 14.09$\pm$0.03 & 11.67$\pm$0.02 & 1 \\
2MASS J09393548$-$2448279$^b$ & T7.5 &--- & 15.70$\pm$0.01 &--- & 17.10$\pm$0.12 & 14.91$\pm$0.03 & 11.64$\pm$0.02 & 2 \\
ULAS J095047.28+011734.3 & T8p & 18.96$\pm$0.06 & 18.05$\pm$0.04 & 18.24$\pm$0.15 & --- & 17.64$\pm$0.18 & 14.51$\pm$0.05 & 4 \\
ULAS J104224.20+121206.8 & T7.5p & 19.58$\pm$0.09 & 18.52$\pm$0.06 & 18.90$\pm$0.12 & --- & 18.23$\pm$0.40 & 15.87$\pm$0.17 & 4 \\
WISE J104245.23$-$384238.3 & T8.5 & --- & 18.81$\pm$0.08 & 19.08$\pm$0.11 & --- & 18.28$\pm$0.30 & 14.56$\pm$0.05 & 6 \\
WISE J115013.85+630241.5 & T8 &--- & 17.73$\pm$0.28 &--- &--- & 16.96$\pm$0.09 & 13.41$\pm$0.03 & 6 \\
WISE J121756.90+162640.8AB & T9 &--- & 17.83$\pm$0.02 & 18.18$\pm$0.05 &--- & 16.55$\pm$0.08 & 13.13$\pm$0.03 & 6 \\
ULAS J123327.45+121952.2 & T4.5 & 19.01$\pm$0.08 & 18.02$\pm$0.04 & 18.22$\pm$0.09 & --- &--- &--- & 11 \\
ULAS J130303.54+001627.7 & T5.5 & 20.11$\pm$0.07 & 19.02$\pm$0.04 & 19.46$\pm$0.11 & 19.94$\pm$0.35 &--- &--- & 13, 12 \\
ULAS J131610.28+075553.0 & sdT6.5 & 20.00$\pm$0.14 & 19.29$\pm$0.12 & --- & --- & $\ge$16.46 & 15.50$\pm$0.10 & 5 \\
ULAS J141623.94+134836.3 & T7.5p & 18.16$\pm$0.03 & 17.26$\pm$0.02 & 17.58$\pm$0.03 & 18.93$\pm$0.24 & 15.99$\pm$0.19 & 12.78$\pm$0.04 & 3 \\
WISEP J142320.86+011638.1 & T8p & 19.69$\pm$0.05 & 18.71$\pm$0.05 & 19.06$\pm$0.08 & 19.89$\pm$0.33 & 17.99$\pm$0.26 & 14.85$\pm$0.07 &  14 \\
WISE J144901.85+114710.9 & T5p & 18.35$\pm$0.04 & 17.36$\pm$0.02 & 17.73$\pm$0.07 & 18.10$\pm$0.15 & 17.07$\pm$0.10 & 14.86$\pm$0.06 & 9 \\
ULAS J150457.66+053800.8 & sdT5.5 & 17.64$\pm$0.02 & 16.59$\pm$0.02 & 17.05$\pm$0.04 & 17.41$\pm$0.09 & 16.00$\pm$0.05 & 14.23$\pm$0.04 & 12, 17 \\
ULAS J151721.12+052929.0 & T8p & 19.57$\pm$0.07 & 18.54$\pm$0.05 & 18.85$\pm$0.15 & --- & 18.17$\pm$0.32 & 15.12$\pm$0.08 & 4 \\
WISE J152305.10+312537.6 & T6.5p &--- & 18.27$\pm$0.07 & 18.69$\pm$0.18 &--- & 17.66$\pm$0.17 & 14.39$\pm$0.04 & 9 \\
WISE J170745.85$-$174452.5 & T5: & --- & 16.25$\pm$0.02 & --- & 16.49$\pm$0.10 & 16.13$\pm$0.07 & 13.65$\pm$0.04 & 9 \\
WISE J174640.78$-$033818.0 & T6 & --- & 17.17$\pm$0.03 & 17.45$\pm$0.04 & --- & 17.87$\pm$0.29 & 14.79$\pm$0.07 & 9 \\
WISE J191359.78+644456.6 & T5: &--- &--- &--- &--- & 17.17$\pm$0.07 & 14.95$\pm$0.04 & 9 \\
WISE J200520.38+542433.9 & sdT8 &--- & 19.64$\pm$0.09 & 19.57$\pm$0.08 &--- & 17.69$\pm$0.14 & 14.87$\pm$0.05 & 10 \\
WISE J201404.13+042408.5 & T6.5p &--- & 18.76$\pm$0.42 &--- &--- & 17.30$\pm$0.17 & 14.96$\pm$0.07 & 9 \\
WISE J212321.92$-$261405.1 & T5.5 & 18.14$\pm$0.03 & 17.04$\pm$0.02 & --- & 17.69$\pm$0.14 & 16.76$\pm$0.11 & 14.45$\pm$0.07 & 9 \\
WISE J213456.73$-$713744.5 & T9p & --- & 19.80$\pm$0.10 & 19.70$\pm$0.15 & --- & 17.61$\pm$0.15 & 13.96$\pm$0.04 & 6 \\
ULAS J223728.91+064220.1 & T6.5p & 19.79$\pm$0.13 & 18.98$\pm$0.12 & 19.23$\pm$0.04 & 19.94$\pm$0.18 & --- &--- & 4  \\
WISE J231939.14$-$184404.4 & T7.5 & 18.61$\pm$0.04 & 17.57$\pm$0.03 & 17.95$\pm$0.05 & 18.55$\pm$0.27 & 16.67$\pm$0.09 & 13.82$\pm$0.04 &  6 \\
WISE J232519.53$-$410535.0 & T9p & --- & 19.53$\pm$0.08 & 19.22$\pm$0.11 & --- & 17.06$\pm$0.11 & 14.11$\pm$0.04 & 6 \\
\hline
\end{tabular}
\begin{list}{}{}
\item[Note.]$^a$All could be classified as sdT subclass. $^b$Unresolved binary candidates (see Section \ref{ssptm} and \citealt{burg08}).  $^c$Reference for $K$-band spectra: 1. \citet{burg02}; 2. \citet{bur06b}; 3. \citet[][SDSS J141624.08+134826.7B]{burn10}; 4. \citet{burn13}; 5. \citet{burn14}; 6. \citet{kirk11}; 7. \citet{lodi07};  8. \citet{loop07};  9. \citet{mace13a}; 10. \citet[][Wolf 1130C]{mace13b}; 11. \citet{maro15}; 12. \citet[][HIP 73786B]{murr11};   13. \citet{pinf08}; 14. \citet[][HIP 70319B]{pinf12}; 15. \citet{pinf14}; 16. \citet{tinn05}; 17. This paper. 
\end{list}
\end{table*}

\begin{figure*}
\begin{center}
  \includegraphics[width=\textwidth]{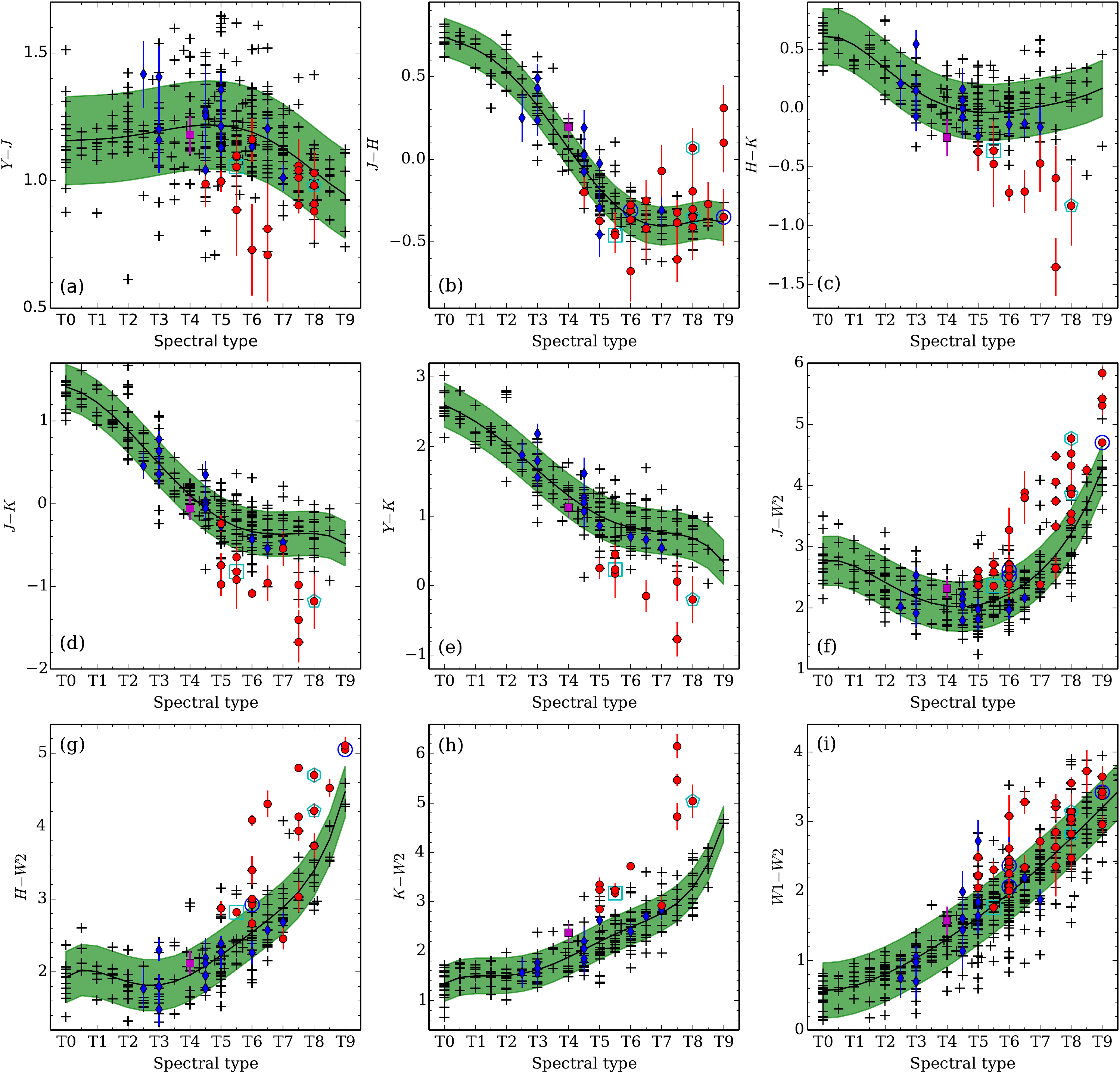}
\caption{Spectral-type versus infrared colour correlations of metal-poor T dwarfs (the red filled circles) compared to T dwarfs (the black crosses). Three T subdwarf companions to bright stars with known metallicity ([Fe/H]) are indicated with the cyan square \citep[HIP 73786B; $-0.3\pm0.1$ dex;][]{murr11}, pentagon \citep[HIP 70319B; $-0.38\pm0.06$ dex;][]{pinf12}, and hexagon \citep[Wolf 1130C; $-0.70\pm0.12$ dex;][]{mace18}. Four close binaries are highlighted with the blue open circles in relevant panels. The blue filled diamonds and a magenta square indicate 14 new T dwarfs and a mildly metal-poor T dwarf identified in this work. The black solid line is the fifth-order polynomial fits to colours of T dwarfs as a function of spectral-type (Table \ref{tsc}). The green shaded areas indicate the rms of these fits. Photometry of known T dwarfs is from UKIDSS and VISTA surveys. }
\label{fsc}
\end{center}
\end{figure*}

\begin{table*}
 \centering
  \caption[]{Coefficients of polynomial fits of UKIDSS--VISTA photometric difference as a function of spectral-types (SpT) in Fig. \ref{fyjhk}, infrared colours as a function of SpT for T0--T9 dwarfs in  Fig. \ref{fsc}, and absolute magnitudes $M_{\rm abs}$ as function of SpT for L and T subdwarfs in Fig. \ref{fsptm}. These fits are defined as  $Colour$ or $M_{\rm abs} = \sum_{i=0}^5 c_i \times ({\rm SpT})^{i}$. In the fits of UKIDSS--VISTA photometric difference (first four rows), SpT = 0 for L0, SpT = 5 for L5, SpT = 10 for T0, SpT = 15 for T5. In fits of SpT--colour and SpT--$M_{\rm abs}$, SpT = 0 for L0, SpT = 5 for L5 dwarfs/subdwarfs, and SpT = 0 for T0, SpT = 5 for T5 subdwarfs. The rms of polynomial fits and $SpT$ ranges are in the last two columns. Coefficients of polynomial fits of $M_{\rm J}$, $M_{\rm H}$, $M_{\rm K}$, and SpT for L subdwarfs are in table 10 of \citetalias{zha18b}. }
 \label{tsc}
  \begin{tabular}{l r r r r r r c l }
\hline
    $Colour/M_{\rm abs}$  & $c_0$~~ & $c_1~~~~$ & $c_2~~~~$ & $c_3~~~~$ & $c_4~~~~$ & $c_5~~~~$ & rms & SpT range  \\
\hline
$Y - Y_{\rm V}$ & 0.0350 & $-$0.0020 &  --- & --- & --- & --- & 0.0952 & L0--T9 \\
$J - J_{\rm V}$ & 0.0263 & $-$0.0026  & --- & --- & --- & --- &  0.0736 & L0--T9 \\
$H - H_{\rm V}$ & 0.0157 & 0.0007  & --- & --- & --- & --- &  0.0986  & L0--T9 \\
$K - Ks_{\rm V}$ & $-$0.0350 & $-$0.0166 & 0.0038 & $-$0.0001 & --- & --- & 0.0845  & L0--T9 \\
$Y-J$ & 1.1562 & 7.6778E$-$3 & $-$6.3168E$-$3 & 5.3137E$-$3 & $-$1.0479E$-$3 & 5.4775E$-$5 & 0.1728 & T0--9 \\
$J-H$ & 0.7387 & $-$7.2589E$-$2 & 2.1773E$-$2 & $-$2.6870E$-$2 & 4.8229E$-$3 & $-$2.4189E$-$4 & 0.1143 & T0--9 \\
$H-K$ & 0.6072 & 6.2116E$-$2 & $-$1.7331E$-$1 & 4.6423E$-$2 & $-$4.7590E$-$3 & 1.7647E$-$4 & 0.2380 & T0--9 \\
$J-K$ & 1.4153 & $-$8.9697E$-$2 & $-$1.0894E$-$1 & 8.1745E$-$3 & 1.6355E$-$3 & $-$1.5167E$-$4 & 0.2684 & T0--9 \\
$Y-K$ & 2.5989 & $-$2.0675E$-$1 & $-$2.1042E$-$2 & $-$1.5787E$-$2 & 4.5339E$-$3 & $-$2.8687E$-$4 & 0.3167 & T0--9 \\
$J-W2$ & 2.7674 & 1.1551E$-$1 & $-$2.6051E$-$1 & 7.1502E$-$2 & $-$7.5839E$-$3 & 3.2506E$-$4 & 0.4022 & T0--9 \\
$H-W2$ & 1.9272 & 3.6843E$-$1 & $-$4.1594E$-$1 & 1.4024E$-$1 & $-$1.8021E$-$2 & 8.2841E$-$4 & 0.3523 & T0--9 \\
$K-W2$ & 1.3432 & 3.9087E$-$1 & $-$3.4851E$-$1 & 1.2996E$-$1 & $-$1.8368E$-$2 & 9.0972E$-$4 & 0.3611 & T0--9 \\
$W1-W2$ & 0.5687 & $-$4.5420E$-$3 & 8.0289E$-$2 & $-$1.1951E$-$2 & 1.1001E$-$3 & $-$3.9575E$-$5 & 0.3973 & T0--9 \\
$M_Y$ & 11.5748 & 0.2443 & --- & --- & --- & --- & 0.1220 & esdL0--3, usdL0--3 \\
$M_Y$ & 12.1660 & 0.3129 & --- & --- & --- & --- & 0.1063 & sdL0--7 \\
$M_{W1}$ & 10.0081 & 0.2610 & --- & --- & --- & --- & 0.1084 & esdL0--7, usdL0--7 \\
$M_{W1}$ & 10.3005 & 0.1787 & --- & --- & --- & --- & 0.1735 & sdL0--7 \\
$M_{W2}$ & 9.7670 & 0.2250 & --- & --- & --- & --- & 0.1213 & esdL0--7, usdL0--7 \\
$M_{W2}$ & 9.9392 & 0.1830 & --- & --- & --- & --- & 0.1647 & sdL0--7 \\
$M_Y$ &  11.2935 & 0.8742 & --- & --- & --- & --- & 0.6426  & sdT5.5--8 \\
$M_J$ & 7.9159 & 1.2294 & --- & --- & --- & --- & 0.8135 & sdT5.5--9 \\
$M_H$ & 9.2711 & 1.0964 & --- & --- & --- & --- & 0.7675 & sdT5.5--9 \\
$M_K$ & 9.1042 & 1.2030 & --- & --- & --- & --- & 0.7876 & sdT5.5--8 \\
$M_{Ks}$ & 8.9844 & 1.1996 & --- & --- & --- & --- & 0.7865 & sdT5.5--8 \\
$M_{W1}$ & 10.5810 & 0.7678 & --- & --- & --- & --- & 0.4584 & sdT5.5--9 \\
$M_{W2}$ & 10.2525 & 0.4232 & --- & --- & --- & --- &  0.2887 & sdT5.5--9 \\
\hline
\end{tabular}
\end{table*}

\begin{figure*}
\begin{center}
  \includegraphics[width=\textwidth]{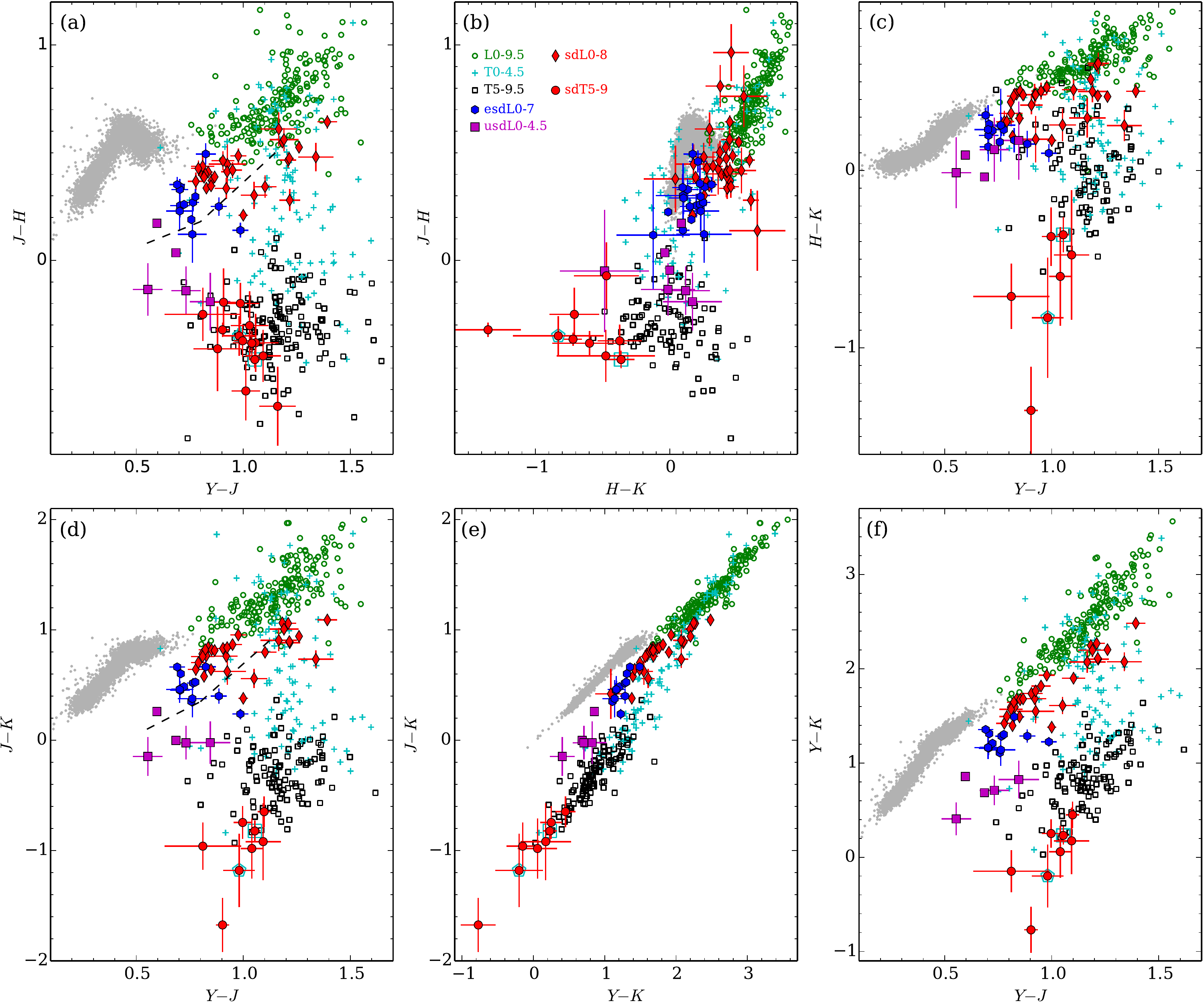}
\caption{Infrared colour--colour diagrams of T5--9 subdwarfs (the red filled circles) compared to other dwarf populations. The grey dots are 5000 point sources selected from a 10 deg$^2$ area of ULAS-SDSS-WISE sky with $14 < J < 16$, which are mostly F, G, K, and M0--4 dwarfs.  HIP 73786B,  HIP 70319B, and  Wolf 1130C are highlighted with the cyan open square, pentagon, and hexagon, respectively. Four close binaries are highlighted with the blue open circles in panels (i--n). Other symbols are indicated in panel (b). The black dashed lines in panels (a, d, m) indicate empirical boundaries between stars and T-BDs \citepalias{zha18b}.}
\label{fcc}
\end{center}
\end{figure*}

\addtocounter{figure}{-1}
\begin{figure*}
\begin{center}
  \includegraphics[width=\textwidth]{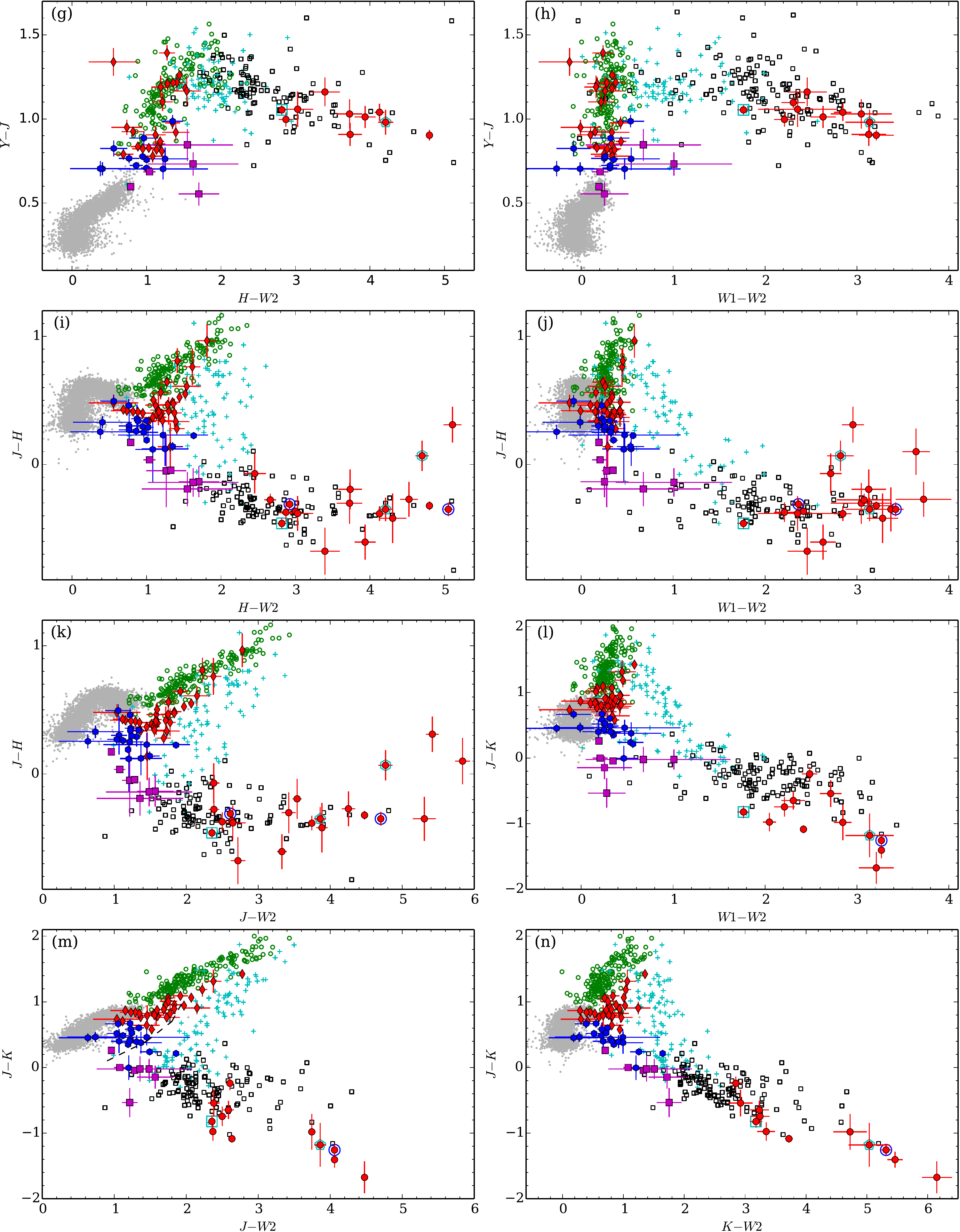}
\caption{Continued.}
\end{center}
\end{figure*}

\section{Population properties of T subdwarfs}
\label{spp} 

\subsection{T subdwarfs in the literature}
\label{sss}
We aimed to find T subdwarfs with UKIDSS and {\sl WISE} surveys (Section \ref{scs}). However, most of our candidates were confirmed as normal T dwarfs. We only identified a mildly metal-poor T dwarf and re-discovered a known T subdwarf (Section \ref{ssc}). A previous search for metal-poor T dwarfs based on UKIDSS and SDSS also found mostly normal T dwarfs \citep{murr11}. 

To further interpret the population properties of T subdwarfs, we searched for them in the literature. We know that metal-poor T dwarfs have suppressed $K$-band flux in their spectra according to those that are wide companions to metal-poor stars. A good example is WISEP J142320.86+011638.1 \citep[HIP 70319B;][]{pinf12}, which is a T8p wide companion to a G1.5V star (HIP 70319) with measured metallicity ([Fe/H] = $-0.38\pm0.06$). Our selection criterion is that the $K$-band peak flux ($F_{\lambda}$) of the object is around or less than half of the T dwarf standard with corresponding spectral subtype when normalized at $J$- and $H$-band. Table \ref{tkmptd} shows the MKO $Y, J, H, K$ and {\sl WISE} $W1, W2$ photometry of 41 T dwarfs with suppressed $K$-band flux that we found in the literature.  

We noticed that only T5+ type metal-poor dwarfs were found by our selection criterion, which is expected by evolutionary models (Section \ref{sdbd}). There are only two early-type T subdwarf candidates reported in the literature. They did not pass our $K$-band spectral selection criterion, thus are not included in our sample. WISE J210529.08$-$623558.7 \citep[WI2105;][]{luhm14} only has $Y, J, H$ band spectrum available that is similar to T dwarf standards. Note this is expected for mildly metal-poor T subdwarfs as we have discussed in Section \ref{spp}. WISE J071121.36$-$573634.2 \citep[WI0711;][]{kell18} has a relative high tangential velocity that consistent with the thick disc. However, its subdwarf nature is not quite clear based on its poor quality NIR spectra. If  WI0711 is accounted as an early-type T subdwarf, the number ratio between T5--9 and T0--4 subdwarfs with $K$-band spectrum is 41. 

The esdT and usdT suclasses have not yet been identified or defined. The unresolved companion of LSR 1610$-$0040 is possibly a esdT6 subdwarf (see Section \ref{shbtz}). WISE J200520.38+542433.9 \citep[Wolf 1130C;][]{mace13b} possibly could be classified into esdT subclass for its significant $Y$-band flux excess \citep[fig. 5]{mace13b}. The kinematics of Wolf 1130A (M2) could be either be halo or thick disc. The optical spectrum of Wolf 1130A fits well with a BT-Settl model \citep{alla14} with [Fe/H] = $-$1.3 \citep[fig. 2;][]{mace18}. Metallicity measurements of Wolf 1130A by absorption lines in the literature are around  [Fe/H] = $-$0.7 and have a large scatter. Note that metallicity measurement of M subdwarfs are not very reliable due to the lack of halo FGK+M subdwarf binaries to calibrate \citep{roja12,zha13,mann15,newt15,pavl15,mont18}.

\begin{table*}
 \centering
  \caption[]{Distance, proper motions (PM), and tangential velocity ($V_{\rm tan}$) of 22 metal-poor T dwarfs. }
\label{tdis}
  \begin{tabular}{l c r r r r r c}
\hline
    Name & $J$  & Distance  & PM$_{\rm RA}$~~ & PM$_{\rm Dec.}$~~ & PM$_{\rm tot}$~~  & $V_{\rm tan}$~~~ & Ref.$^a$   \\
& & (pc)~~~ &  (mas yr$^{-1}$) &   (mas yr$^{-1}$) &   (mas yr$^{-1}$) & (km s$^{-1}$)  & \\
\hline
WISE J023318.05+303030.5$^b$  & 16.86$\pm$0.34 & 31.95$^{+4.95}_{-3.78}$  & $-$134.6$\pm$2.3 & $-$29.5$\pm$2.2 & 137.80$\pm$3.20 & 20.87$^{+3.27}_{-2.52}$ & 7\\
WISE J052844.51$-$330823.9 & 16.90$\pm$0.07 & 20.24$^{+1.74}_{-1.48}$  & 1.0$\pm$1.2 & $-$19.5$\pm$1.6 & 19.50$\pm$2.00 & 1.87$^{+0.25}_{-0.24}$ & 7\\
WISE J061213.85$-$303612.5AB & 16.65$\pm$0.03 & 24.33$^{+0.99}_{-0.91}$  & $-$120.3$\pm$0.7 & $-$258.5$\pm$1.7 & 285.10$\pm$1.80 & 32.88$^{+1.35}_{-1.25}$& 7\\
2MASS J07290002$-$3954043 & 15.64$\pm$0.08 & 7.92$^{+0.56}_{-0.49}$  & $-$566.6$\pm$5.3 & 1643.4$\pm$5.5 & 1738.30$\pm$7.60 & 65.24$^{+4.60}_{-4.03}$& 4\\
WISE J083337.83+005214.2 & 20.28$\pm$0.14 & 12.11$^{+0.70}_{-0.63}$  & 790.1$\pm$2.8 & $-$1590.7$\pm$2.5 & 1776.10$\pm$3.80 & 101.92$^{+5.88}_{-5.27}$& 8\\
WISE J083641.10$-$185947.0 & 19.10$\pm$0.12 & 24.51$^{+2.37}_{-1.99}$  & $-$47.6$\pm$1.3 & $-$150.5$\pm$1.2 & 157.80$\pm$1.80 & 18.33$^{+1.79}_{-1.50}$& 8\\
ULAS J090116.23$-$030635.0 & 17.89$\pm$0.04 & 15.97$^{+0.69}_{-0.64}$  & $-$38.6$\pm$2.3 & $-$261.2$\pm$2.8 & 264.00$\pm$3.60 & 19.99$^{+0.91}_{-0.84}$& 3\\
2MASS J09373487+2931409 & 14.29$\pm$0.03 & 6.14$^{+0.15}_{-0.14}$  & $-$1329.8$\pm$7.1 & 997.0$\pm$7.1 & 1662.00$\pm$7.10 & 48.39$^{+1.21}_{-1.15}$& 1\\
2MASS J093935.93$-$244838.9$^b$ & 15.70$\pm$0.01 & 5.34$^{+0.13}_{-0.13}$  &  573.4$\pm$2.3 & $-$1044.7$\pm$2.5 & 1191.70$\pm$3.40 & 30.16$^{+0.76}_{-0.73}$& 2\\
ULAS J095047.28+011734.3 & 18.05$\pm$0.04 & 17.27$^{+0.71}_{-0.66}$  & 241.6$\pm$0.7 & $-$360.7$\pm$1.1 & 434.10$\pm$1.30 & 35.54$^{+1.47}_{-1.36}$& 7\\
WISE J104245.23$-$384238.3 & 18.81$\pm$0.08 & 15.29$^{+0.84}_{-0.76}$  & $-$75.1$\pm$6.2 & 56.0$\pm$6.2 & 93.70$\pm$6.20 & 6.79$^{+0.58}_{-0.56}$& 5\\
WISE J115013.85+630241.5 & 17.73$\pm$0.28 & 8.03$^{+0.20}_{-0.19}$  & 410.4$\pm$1.8 & $-$539.7$\pm$1.2 & 678.00$\pm$2.20 & 25.81$^{+0.64}_{-0.61}$& 7\\
WISE J121756.90+162640.8AB & 17.83$\pm$0.02 & 9.58$^{+0.45}_{-0.41}$  & 758.1$\pm$1.7 & $-$1252.2$\pm$1.5 & 1463.80$\pm$2.30 & 66.46$^{+3.13}_{-2.86}$& 7\\
ULAS J141623.94+134836.3 & 17.26$\pm$0.02 & 9.30$^{+0.03}_{-0.03}$  & 85.7$\pm$0.7 & 129.1$\pm$0.5 & 154.93$\pm$0.84 & 6.83$^{+0.04}_{-0.04}$& 6\\
WISEP J142320.86+011638.1 & 18.71$\pm$0.05 & 17.45$^{+0.03}_{-0.03}$  & 223.5$\pm$0.1 & $-$478.3$\pm$0.1 & 527.98$\pm$0.10 & 43.68$^{+0.08}_{-0.08}$& 6\\
ULAS J150457.66+053800.8 & 16.59$\pm$0.02 & 19.01$^{+0.04}_{-0.04}$  & $-$607.7$\pm$0.1 & $-$506.5$\pm$0.1 & 791.07$\pm$0.16 & 71.29$^{+0.14}_{-0.14}$& 6\\
ULAS J151721.12+052929.0 & 19.57$\pm$0.07 & 23.75$^{+2.63}_{-2.15}$  & $-$55.5$\pm$2.3 & 194.9$\pm$2.2 & 202.60$\pm$3.20 & 22.81$^{+2.55}_{-2.10}$& 7\\
WISE J152305.10+312537.6 & 18.27$\pm$0.07 & 16.29$^{+1.07}_{-0.95}$  & 103.9$\pm$1.7 & $-$509.2$\pm$1.7 & 519.70$\pm$2.40 & 40.11$^{+2.65}_{-2.35}$& 7\\
WISE J200520.38+542433.9 & 18.64$\pm$0.09 & 16.56$^{+0.01}_{-0.01}$  & $-$1159.5$\pm$0.1 & $-$903.6$\pm$0.1 & 1469.97$\pm$0.11 & 115.38$^{+0.07}_{-0.07}$& 6\\
WISE J213456.73$-$713744.5 & 19.80$\pm$0.10 & 9.12$^{+0.32}_{-0.30}$  & $-$223.2$\pm$6.2 & 1363.2$\pm$6.2 & 1381.40$\pm$6.20 & 59.69$^{+2.10}_{-1.97}$& 5\\
WISE J231939.14$-$184404.4 & 17.57$\pm$0.03 & 10.70$^{+0.32}_{-0.30}$  & 65.5$\pm$1.6 & 138.8$\pm$1.6 & 153.40$\pm$27.50 & 7.78$^{+1.41}_{-1.41}$& 7\\
WISE J232519.53$-$410535.0 & 19.53$\pm$0.08 & 9.23$^{+0.33}_{-0.30}$  & $-$17.5$\pm$6.7 & 836.8$\pm$6.7 & 837.00$\pm$6.70 & 36.60$^{+1.33}_{-1.24}$& 5\\
\hline
\end{tabular}
\begin{list}{}{}
\item[]$^a$Reference of parallaxes: 1. \citet{vrba04}; 2. \citet{burg08}; 3. \citet{maro10}; 4. \citet{fah12}; 5. \citet{tinn14}; 6. \citet[][from their primary companions]{gaia18}; 7. \citet{kirk19}. $^b$Unresolved binary candidates.  
\end{list}
\end{table*}

\subsection{Spectral-type--colour correlations}
Fig. \ref{fsc} shows the correlations between spectral-types and nine infrared colours of T subdwarfs. Note most of these objects do not have photometry in all six infrared bands (Table \ref{tkmptd}). To provide a reference, we also plotted a sample of known field T dwarfs from this work and the literature \citep{burg02,chiu06,lodi07,pinf08,burn10b,burn13,albe11,kirk11,mace13a,maro15,card15,skrz16}. NIR photometry of field T dwarfs are from LAS, VHS, and VIKING surveys. VHS photometry are used when LAS photometry is not available. VIKING photometry was used when both LAS and VHS are not available. We fitted the correlations between spectral-type and nine infrared colours of field T dwarfs by fifth-order polynomial functions. The coefficients of these fitting are listed in Table \ref{tsc}. 

All of these T subdwarfs have $Y-J$ colour bluer than the average of field T dwarfs. ULAS J131610.28+075553.0 \citep[UL1316;][]{burn14} and ULAS J012855.07+063357.0 \citep[UL0128;][]{burn13} have very blue $Y-J$ colour. Note that Wolf 1130C should also have very blue $Y-J$ colour but have no $Y$-band photometry available. The $J-H$ colour of T subdwarfs seems similar to that of field T dwarfs, but more metal-poor T subdwarfs tend to have redder $J-H$ colour (e.g. Wolf 1130C). WISE J232519.53$-$410535.0 \citep[WI2325;][]{kirk11} has the reddest $J-H$ colour in Fig. \ref{fsc} (b). However, most T subdwarfs have similar  $J-H$ colour to T dwarfs, and some also have bluer $J-H$ colour than T dwarfs \citep[e.g. HIP 73786B;][]{murr11}. It seems that $J-H$ colour is not monotonous as a function of metallicity between $-1 \leq$ [Fe/H] $\leq 0$.  $J-H$ colour outliers were also discussed in the literature \citep{mace13b,logs18}. 

T subdwarfs have significant bluer $Y-K, H-K$, and $J-K$ colours than field T dwarfs (Figs \ref{fsc}c--e). This is expected, as $K$-band flux suppression is a signature of low metallicity caused by enhanced collision-induced H2 absorption \citep[CIA H$_2$; e.g.][]{saum12}. ULAS J141623.94+134836.3 3. \citep[UL1416;][]{burn10} has the bluest $Y-K, H-K$, and $J-K$ colours. HIP 70319B and 2MASS J09393548$-$2448279 \citep[2M0939;][]{bur06b} also have very blue $J-K$ colour. Note UL1316 and Wolf 1130C are more metal-poor, but are not shown in Figs \ref{fsc}(c--e) because of the lack of $K$-band photometry. 

The $J$- and $H$-band is the ideal wavelength for the spectral typing of T subdwarfs. The $J-H$ colour is similar for both T dwarfs and mildly metal-poor T subdwarfs. Therefore, $J$- and $H$-band spectra can be used to assign spectral subtype of T subdwarfs, by comparing their $J$- and $H$-band spectra to that of T dwarf standards. However, this becomes a bit more complicated for more metal-poor T subdwarfs, which have redder $J-H$ colour. For example, the sdT8-type Wolf 1130C has much weaker methane absorptions around 1300 and 1650 nm than field T8 dwarfs \citep[fig. 5;][]{mace13b}. Meanwhile, $Y$- and $K$-band spectra is sensitive to metallicity differential and can be used to scale metallicity and define subclass of T subdwarfs. 

T subdwarfs have redder $J-W2$ and $H-W2$ colours than T dwarfs (Figs \ref{fsc}f and g). T subdwarfs with the most extreme $J-W2$ colour also have strongest suppression of $K$-band flux, and among the most metal-poor T subdwarfs, such as WISE J061407.49+391235.9 \citep[WI0614;][]{kirk11}, WISE J152305.10+312537.6 9. \citep[WI1523;][]{mace13a}, UL1416, Wolf 1130C, and WI2325. The more metal-poor the redder the $J-W2$ colour. 

T subdwarfs have redder $W1-W2$ colours than T dwarfs. However, these T subdwarfs, with reddest $J-W2$ colour, do not necessarily have also the reddest $W1-W2$ colour. Only WI1523 have very red $J-W2$ and $W1-W2$ colours. The {\sl WISE} $W1$ photometry of some bluer T subdwarfs could be contaminated by bluer background sources for its low spatial resolution. For example, UL1316 has $W1-W2 \approx 1.0$, which is bluer than any other T5+ subdwarfs. UL1316 was next to a background source when it was observed by {\sl WISE}. T subdwarfs have extremely red $K-W2$ colours compared to T dwarfs, as they have suppressed $K$-band flux and red $W1-W2$ colours. UL1416, 2M0939, HIP 70319B have the reddest $K-W2$ colour in Fig. \ref{fsc} (h), while some more metal-poor ones were not shown due to the lack of corresponding photometry (e.g. Wolf 1130C). 

\subsection{Infrared colour--colour plots}
Photometric colours are easier to obtain than spectra, thus are widely used in the selection and study of celestial populations. Fig. \ref{fcc} shows various combinations of infrared colour--colour plots of stellar and substellar populations of sdT, sdL, esdL, usdL subdwarfs, and  F--M4, L, T dwarfs. Three T subdwarfs in binary systems with known metallicity, HIP 73786B,  HIP 70319B, and  Wolf 1130C, are highlighted. Note some T subdwarfs are not shown in colour--colour plots involve $Y$ and $K$ photometry since they are not available (see Table \ref{tkmptd}). 

We can see that both sdL and sdT5+ subdwarfs have offsets from L and T5+ dwarfs in all of these colour--colour plots in Fig. \ref{fcc}. L subdwarfs are bluer than L dwarfs in all of these colour--colour plots, and the more metal-poor the bluer they are. While sdTs generally have bluer $YJHK$ colours (except $J-H$) and redder colours involving $W2$ photometry. Most of these known T subdwarfs in Table \ref{tkmptd} were found in {\sl WISE}. This is probably because they are relatively bright in $W2$-band and have red $W1-W2$ colour. The T extreme subdwarf (esdT) or ultrasubdwarf (usdT) subclasses could be fainter in $K$ band, brighter in $Y$- and $W2$-bands, and have more extreme colours in these plots. 

The black dashed lines in Fig. \ref{fcc} (a, d, m) indicates empirical boundaries between VLMS and T-BDs \citepalias{zha18b}. A gap is revealed between early sdL and sdT5+ subdwarfs due to the STZ (Section \ref{sebd}).

\begin{figure*}
\begin{center}
  \includegraphics[width=\textwidth]{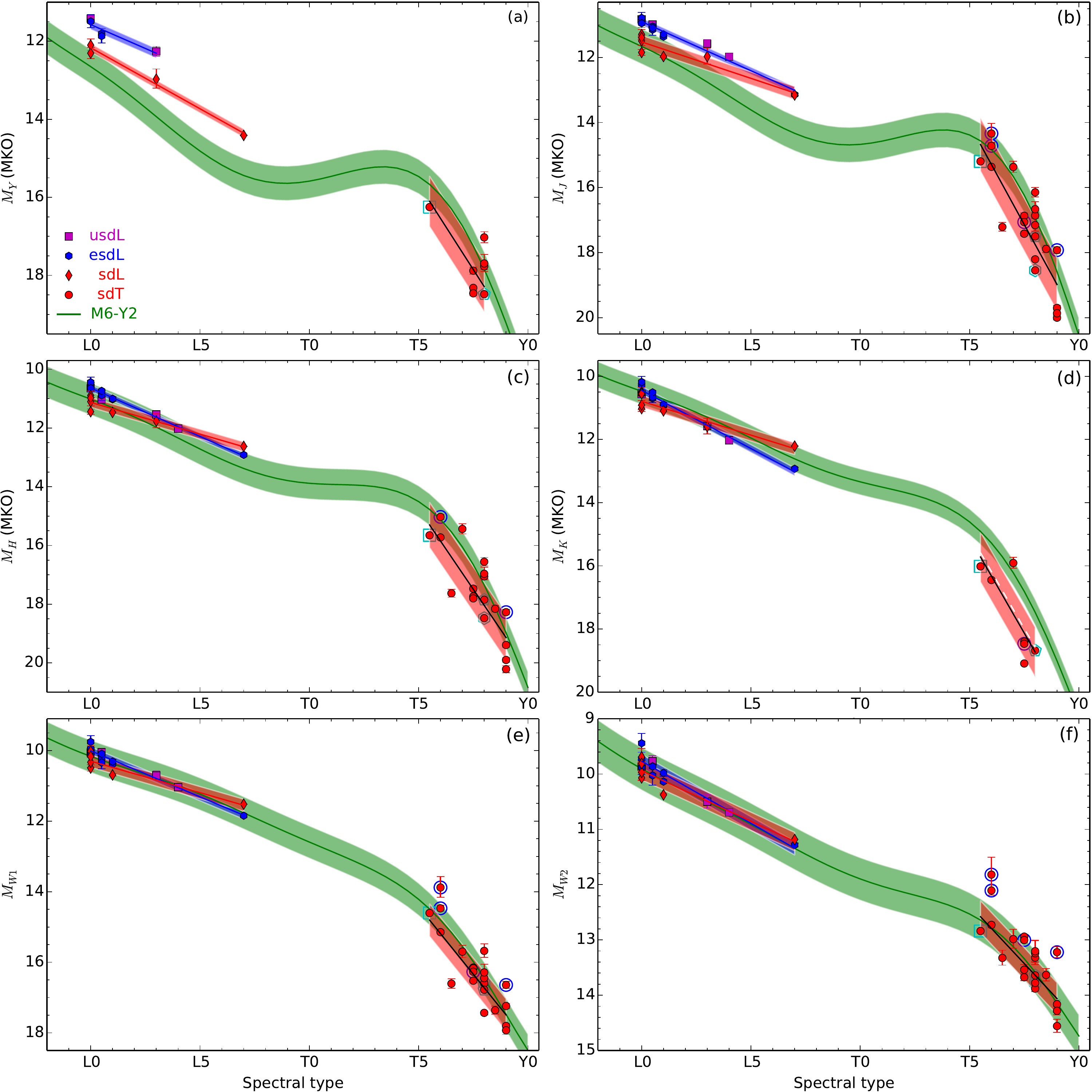}
\caption{Spectral-type and absolute magnitude correlations of L and T subdwarfs compared to that of L and T dwarfs. The black lines are 1st order polynomial fits for T5+ subdwarfs. The white dashed line in panel (d) is for $Ks$-band. The red and blue lines are first-order polynomial fits for L subdwarfs \citepalias{zha18b}. The green lines are seventh-order polynomial fits for field M6--Y2 dwarfs with known parallaxes \citep{dupu12,dupu13}. HIP 73786B,  HIP 70319B, and  Wolf 1130C are highlighted with the cyan open square, pentagon, and hexagon, respectively. Objects highlighted with the blue circles are two known binaries and two binary candidates and not used for the fitting. The shaded areas indicate the rms of these fits. Coefficients of these polynomial fits are in Table \ref{tsc}. Reference for parallaxes of T subdwarfs are indicated in Table \ref{tdis}. {\sl Gaia} DR2 parallaxes of 20 L subdwarfs are listed in table 9 of \citetalias{zha18b}.}
\label{fsptm}
\end{center}
\end{figure*}

\subsection{Spectral-type--absolute magnitude correlations}
\label{ssptm}
Spectral-type--absolute magnitude correlation is often used to estimate spectroscopic distance of cool and ultracool dwarfs, and reveals physical properties of different populations. The absolute magnitude or luminosity of an ultracool dwarf is mainly depending on its radius and $T_{\rm eff}$. Ultracool dwarfs with the same mass and different metallicity also have different $T_{\rm eff}$. Meanwhile, spectral-type is based on empirical classification of observed spectra. Therefore, the spectral-type--absolute magnitude correlations for objects with different spectral-type and metallicity are also different. The difference of their spectral-type--absolute magnitude correlation also unveils the difference of their physical properties. 

Table \ref{tdis} shows the distances and proper motions of 22 T subdwarfs with measured parallaxes in the literature \citep{vrba04,burg08,maro10,fah12,tinn14,gaia18,kirk19}. We calculated their absolute magnitudes in MKO $Y, J, H, K$, and {\sl WISE} $W1, W2$ band. Fig. \ref{fsptm} shows the polynomial fits of spectral-type--absolute magnitude correlations of these T subdwarfs compared to that of M6--Y2 dwarfs and L0--7 subdwarfs. Distances of L subdwarfs are from {\sl Gaia} DR2 \citepalias{zha18b}. Distances of M6--Y2 dwarfs are from \citet{dupu12} and \citet{dupu13}. Four T subdwarfs were excluded in our polynomial fits. Because WISE J061213.85$-$303612.5AB and WISE J121756.90+162640.8AB \citep{kirk11} are listed as binaries in \citet{kirk19}. WISE J023318.05+303030.5 \citep{mace13a} has the brightest  $M_{J}$, $M_{W1}$, and $M_{W2}$, thus likely is an unresolved binary (see panels e--f of Fig. \ref{fsptm}). 2M0939 is an unresolved binary candidate \citep{burg08}.  

Fig.  \ref{fsptm} shows that absolute magnitudes have a steep decline from T5 as a function of the spectral-type. $M_{Y, J, H, K, W1}$ of T5+ subdwarfs are fainter than T5 dwarfs by 0.5--1.5 mag, and $M_{K}$ is the most distinct.  $M_{W2}$ of T5+ subdwarfs and dwarfs are similar. Note the metallicity of these T5+ subdwarfs is likely between $-1 \la$ [Fe/H] $\la -0.3$ (by these in wide binary systems with known metallicity in Fig. \ref{fsc}), which is equivalent to the metallicity range of sdL subclass \citepalias{zha17a}. Absolute magnitudes of esdT5+ and usdT5+ at $Y, J, H, K$, and $W1$ band would likely follow the trend from T5+ dwarfs to subdwarfs and fainter than sdT5+ subclass.  

Unlike T5+ subdwarfs, L0-7 subclasses at lower metallicity (esdL and usdL) have brighter absolute magnitude from the optical $G$ to NIR $H$ bands. Only $M_{K}$ is slightly fainter for esdL/usdL3-7 subdwarfs. $M_{W1}$ and $M_{W2}$ at L0--7 subtype do not react much to metallicity differential. Note the correlations between spectral-type and $M_{G, BP, RP, i, z, y}$ for L subdwarfs are presented in fig. 22 of \citetalias{zha18b}.  

The different reaction of spectral-type--absolute magnitude correlation to metallicity differential for L0--7 and T5+ subdwarfs are related to the hydrogen burning transition zone or the STZ (Section \ref{shbtz}). T5+ subdwarfs are D-BDs (the majority of BDs) without hydrogen fusion. They are more massive than T5+ dwarfs with the same subtype, but they have much longer cooling time ($\ga$ 8 Gyr) and become fainter than T5+ dwarfs. L subdwarfs are also more massive than L dwarfs with the same subtype, as the HBMM is higher at lower metallicity (fig.9; \citetalias{zha17b}). L0--7 subdwarfs are 200--400 K hotter than L dwarfs (fig. 4; \citetalias{zha18a}). Early-type L subdwarfs are VLMS with sustained hydrogen fusion. While L3--7 subdwarfs are massive T-BDs with about 50--99 per cent of their energy provided by unsteady hydrogen fusion. L subdwarfs have lower opacity and emit relatively more energy at short wavelengths. In consequence, L0--7 subdwarfs have brighter $M_{G, RP, Y, J, H}$ than L dwarfs that are composed of D-BDs, T-BDs, and some VLMS. 

Without hydrogen fusion, metal-poor D-BDs have migrated to the T5 subdwarf region at the age of $\ga$ 8 Gyr. Meanwhile, T-BDs in the STZ have low-rate hydrogen fusion in their cores that is contributing 1--99 per cent of their luminosity (depending on their mass) with the dissipation of their initial thermal energy \citep{burr11}. Consequently, the spectral-type range of metal-poor T-BDs is stretched to $\sim$L3 -- T4 after $\ga$ 8 Gyr. 

The spectral-type distributions of thin-disc T-BDs and D-BDs (dwarfs) are more complicated. Depending on the age, spectral-types of VLMS are between late-type M and $\sim$L3, T-BDs are between late-type M and T5, D-BDs are between late M and Y types. Therefore, late-type M--L3 dwarfs are composed of VLMS, T-BDs, and D-BDs. L3--T5 dwarfs are composed of T-BDs and D-BDs. T5+ dwarfs are all D-BDs.

\subsection{Infrared HR diagrams}
The Hertzsprung--Russell diagram is the most important diagram in the study of populations of the Milky Way. Fig. \ref{fhrd} shows various infrared HR diagrams of T5+ subdwarfs compared to those of L/T dwarfs, and sdL, esdL, usdL subdwarfs. We highlighted five L subdwarfs that are T-BDs, three T subdwarf that are wide companions to stars with known metallicity, and four close T subdwarf binaries in Fig \ref{fhrd}. L and T dwarfs that appear as a zigzag sequence on these HR diagrams involve NIR colours ($J-H$, $J-K$) due to their complex atmospheres. This is different from main-sequence stars. The zigzag sequence also appears on the $J-W2$ HR diagrams, but less obvious. It is barely visible on the $H-W2$  HR diagrams, and absent for $K-W2$ and $W1-W2$ colours. 

L subdwarfs have bluer $YJHK$ colours and brighter absolute magnitudes ($M_G$ to $M_H$) than L dwarfs and appear on the upper left-hand side of the L dwarf sequence on NIR HR diagrams (Figs \ref{fhrd} a, b, d, e, g, h, p, q). The location of L subdwarf sequences from the sdL, to esdL and usdL subclass (towards lower metalliciy) are further and further away from L dwarf sequence. We plotted 10 Gyr isochrones of  0.12--0.083 M$_{\sun}$ objects with [M/H] = $-$1 ([Fe/H] $= -$1.3) and [M/H] = $-$2 ([Fe/H] $= -$2.3) from \citet{bara97} in Figs \ref{fhrd} (a, b, d, e, g, h). We also marked the SHBMM at [Fe/H] $= -$1.3 (0.083 M$_{\sun}$) and [Fe/H] $= -$2.3 (0.0875 M$_{\sun}$) on these isochrones \citepalias{zha17b}. The faintest point of these isochrones is at 0.083 M$_{\sun}$ and [Fe/H] $= -$2.3, corresponding to a late usdL or early-type-usdT T-BD according to fig. 9 of \citetalias{zha17b}.  

T5+ subdwarfs have fainter $M_{J,H,K}$ than T5+ dwarfs with the same subtype, thus appear lower than T dwarfs in panels  of Fig. \ref{fhrd}. They also have bluer $J-K$ colour and appear shifted to the lower left from T5+ dwarf sequence in panels (b, e, h). T5+ subdwarf sequence shifted toward lower right from T5+ dwarf sequence in panels (c, f, i, j, k, l), as they have fainter $M_{J, H, K, W1}$ and redder $Y/J/H/K/W1 - W2$ colours. T5+ subdwarf sequence appears to the right of T5+ dwarf sequence in panels (m, n, o, r), for their redder $J/H/K/W1 - W2$ colours and similar $M_{W2}$.  

The region between two turns of the zigzag sequence of L and T dwarfs on the HR diagram is referred to as the L/T transition (e.g. Figs \ref{fhrd} a--c). The sdL and sdT sequence does not follow a zigzag sequence as the L and T dwarfs in HR diagrams involve $J-H, J-K, J-W2$ colours (Figs \ref{fhrd} a--e, g, h, j, m, p, q ). First, because L and T subdwarfs have less molecules in their atmospheres than L and T dwarfs due to lower metallicity. Secondly, metal-poor D-BDs have longer cooling time ($\ga$ 8 Gyr), and have left the L and T0--4.5-type sequence and become T5+ subdwarfs. Thirdly, the L and T subdwarf sequence was stretched by the STZ. The stretched $M_{J,H,K,W1,W2}$ ranges of the `sd' subclass T-BDs in the STZ are indicated with the grey stripes in panels (a, d, g, j, m) of Fig. \ref{fhrd}. Note that the absolute magnitude range of T-BDs of the esd and usd subclasses should be broader than that of the sd subclass. 

The majority of metal-poor BDs below the SHBMM are D-BDs with T5+ types. Only a small fraction of BDs is T-BDs that populated between $11.5 \la M_J \la 15.0$. The number density of T5+ subdwarfs is much higher than that of L3--T4.5 subdwarfs. There are 41 known T5+ subdwarfs in Table \ref{tkmptd}. Meanwhile, there are 39 known sdL subdwarfs in table 1 of \citetalias{zha18b}. The average distance of known L subdwarfs in {\sl Gaia} DR2 (fig. 23; \citetalias{zha18b}) are about five times larger than these T5+ subdwarfs in Table \ref{tdis}. Thus, the discovery volume of known sdL subdwarfs is about 125 times larger than that of known T5+ subdwarfs. This implies that the number density of sdT5+ subdwarfs in the solar neighbourhood is about a 100 times higher than that of sdL subwarfs. There are much more T5+ subdwarfs awaiting to be discovered in the solar neighbourhood.

\begin{figure*}
\begin{center}
  \includegraphics[width=\textwidth]{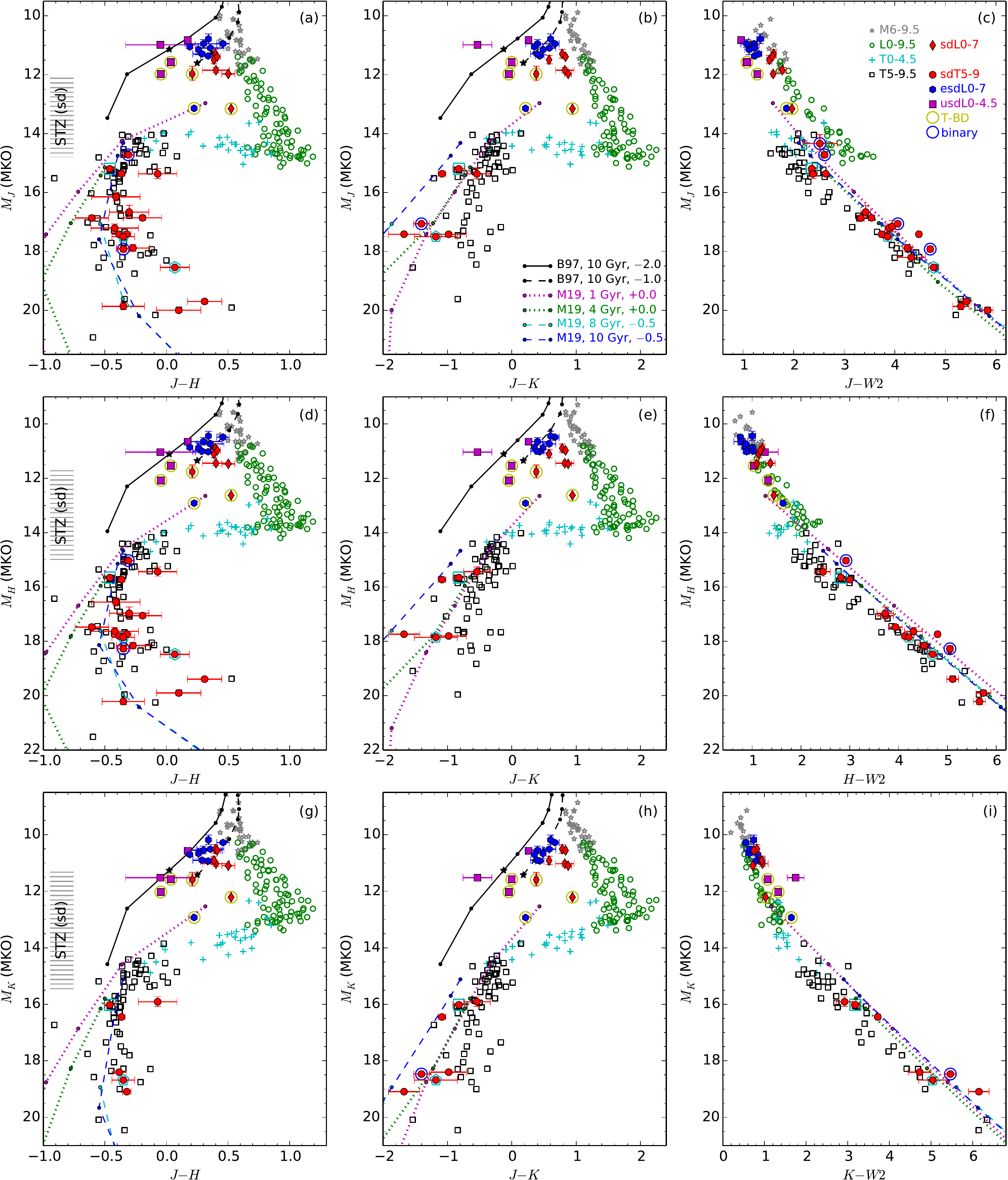}
\caption{Infrared HR diagrams for populations of L, T subdwarfs, and M6+, L, T dwarfs (indicated in panel c). HIP 73786B,  HIP 70319B, and  Wolf 1130C are highlighted with the cyan open square, pentagon, and hexagon, respectively. Objects highlighted with the blue circles are two known binaries and two binary candidates. Five metal-poor T-BDs are highlighted with the yellow circles. The stretched $M_{J,H,K,W1,W2}$ ranges of `sd' subclass T-BDs in the STZ are indicated with the grey stripes in panels (a, d, g, j, m). The 10 Gyr isochrones of 0.12--0.083 M$_{\sun}$ objects are plotted as the dashed ([Fe/H] = $-$1.0, i.e. [Fe/H] $\approx -$1.3) and solid ([Fe/H] = $-$2.0, i.e. [Fe/H] $\approx -$2.3) lines \citep{bara97}. The SHBMM at [Fe/H] = $-$1.3 (0.083 M$_{\sun}$) and [Fe/H] = $-$2.3 (0.0875 M$_{\sun}$) is marked with the two black five-pointed stars \citepalias{zha17b}. The isochrones with [M/H] = 0.0 (dotted) at 1 (magenta) and 4 (green) Gyr, and [M/H] = $-0.5$ ([Fe/H] $\approx -0.7$; dashed) at 8 (cyan) and 10 (blue) Gyr are from Marley et al. (2019, in prep).  
}
\label{fhrd}
\end{center}
\end{figure*}

\addtocounter{figure}{-1}
\begin{figure*}
\begin{center}
  \includegraphics[width=\textwidth]{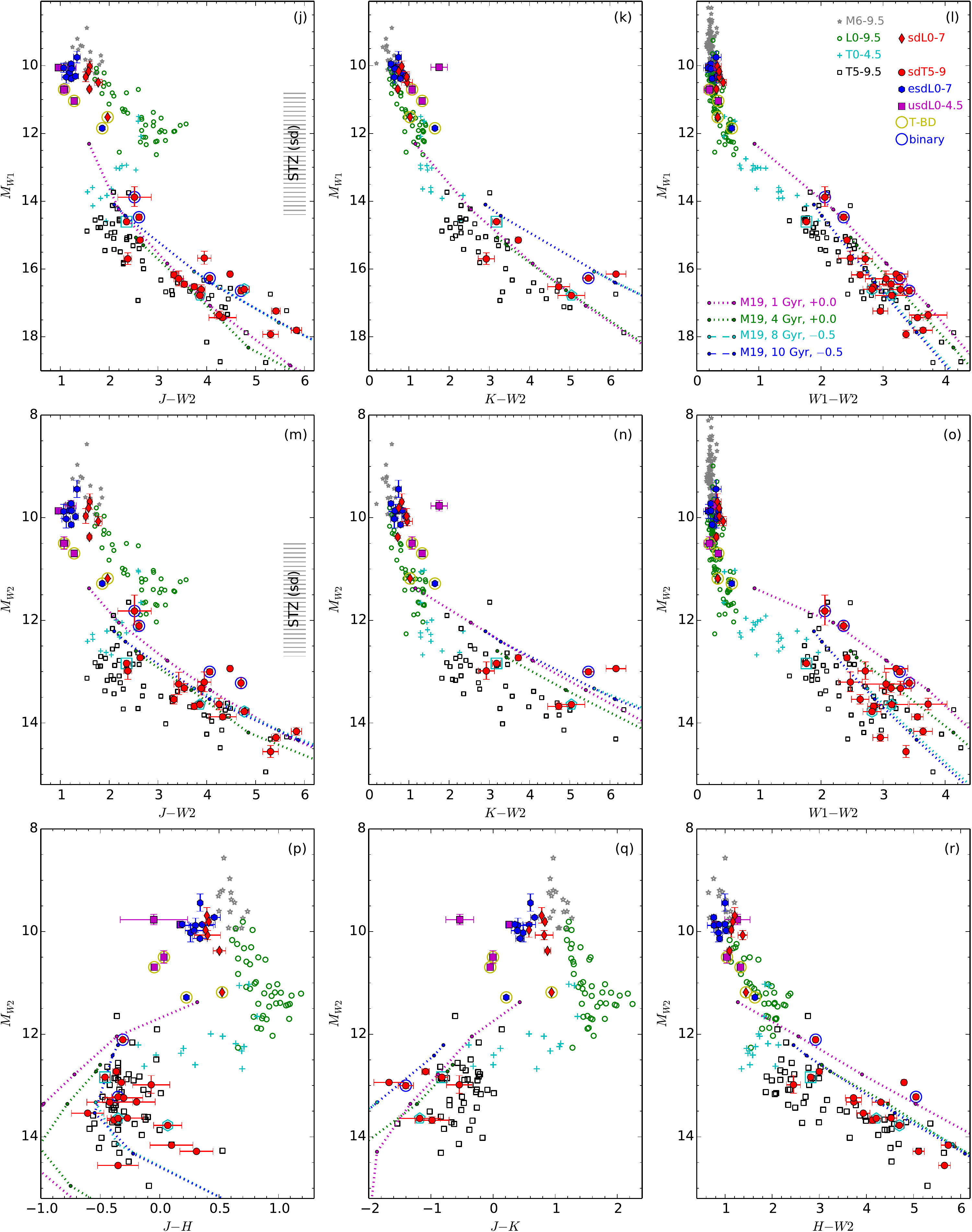}
\caption{Continued.}
\label{fhrd2}
\end{center}
\end{figure*}

\section{Discovery capability of future surveys}
\label{sfss}
Most of the T5--9 subdwarfs have temperature between 1000 and 500 K (Fig. \ref{fiso}). They emit most of their flux beyond 900 nm wavelength and are extremely faint (Fig. \ref{fsptm}). Future deep infrared sky surveys have capability to identify a large number of T5+ subdwarfs. These surveys include {\sl Euclid} \citep{laur11}, {\sl Wide-Field InfraRed Survey Telescope} \citep[{\sl WFIRST};][]{sper15},  {\sl CatWISE}, {\sl Chinese Space Station Optical Survey} ({\sl CSS-OS}), and  Large Synoptic Survey Telescope \citep[LSST;][]{lsst17}. Note, {\sl Euclid} and {\sl WFIRST} photometric colours of solar metallicity field 
BDs are discussed in \citet{holw18}.   

We estimated the numbers of T5--9 subdwarfs that could be detected in these surveys based on their survey depth and coverage, and the absolute magnitude of T5--9 subdwarfs. T9 subdwarfs are fainter than T5 subdwarfs by several magnitudes in the infrared. A sky survey can detect T5 subdwarfs out to a further distance than T9 subdwarfs at a limiting magnitude. We estimated the maximum distance of T7 subdwarfs at the limiting magnitude of each survey, and used it as an average maximum distance for T5--9 subdwarfs. The maximum distance of T7 subdwarfs in a sky survey is calculated from the difference between the absolute magnitude of T7 subdwarfs and the limiting magnitude of the sky survey. $M_{Y, J, H, K, W1, W2}$ of T7 subdwarfs are from Fig. \ref{fsptm} and Table \ref{tsc}. The spectral-type--$M_{z, y}$ correlation is not available for T subdwarfs. Therefore, we used the absolute magnitudes of an sdT7.5 subdwarf (2M0939) in Pan-STARRS \citep{cham16} $z_{\rm P1}$ and $y_{\rm P1}$ band to estimate the average maximum distance of T5--9 subdwarfs at $z$ and $y$ band limiting magnitudes of {\sl CSS-OS} and LSST. Note that 2M0939 is a candidate of unresolved binary \citep{burg08}. If 2M0939 is an equal-mass binary, its $M_{z, y}$ should be 0.75 mag brighter than a single sdT7.5 subdwarf, and should be similar to that of an sdT6.5--7 subdwarf. The binarity of 2M0939 does not have a big impact on the estimated detection depths of T subdwarf for {\sl CSS-OS} and LSST. 

We use the number density of stars, fractions of thick disc and halo stars (assume it is the same for BDs), and star/T dwarf ratio in the solar neighbourhood, to estimate the number density of T5--9 subdwarfs of the thick disc and halo. We found 1335 objects within 20 pc (parallax $\geq$ 50.0 and parallax/parallax\_error $\geq$ 5) of the Sun in half of the sky towards Galactic caps ($|b| > 30$ deg) in the {\sl Gaia} DR2 \citep{gaia18}. Then we know that the number density of stars in the solar neighbourhood is around 0.079677 pc$^{-3}$. We estimated the star/T dwarf ratio to be around 10. As there are at least 211 star and 23 T dwarfs within 8 pc of the Sun \citep{kir12}. Note metal-poor D-BDs have become T5--9 and Y subdwarfs. We assumed that the ratio between T5--9 and Y subdwarfs is similar to the ratio between T- and L + Y-type D-BDs. We also know the fractions of thick and halo stars in the solar neighbourhood is about 0.07 and 0.006, respectively \citep{redd06}. 

\begin{figure}
\begin{center}
  \includegraphics[width=\columnwidth]{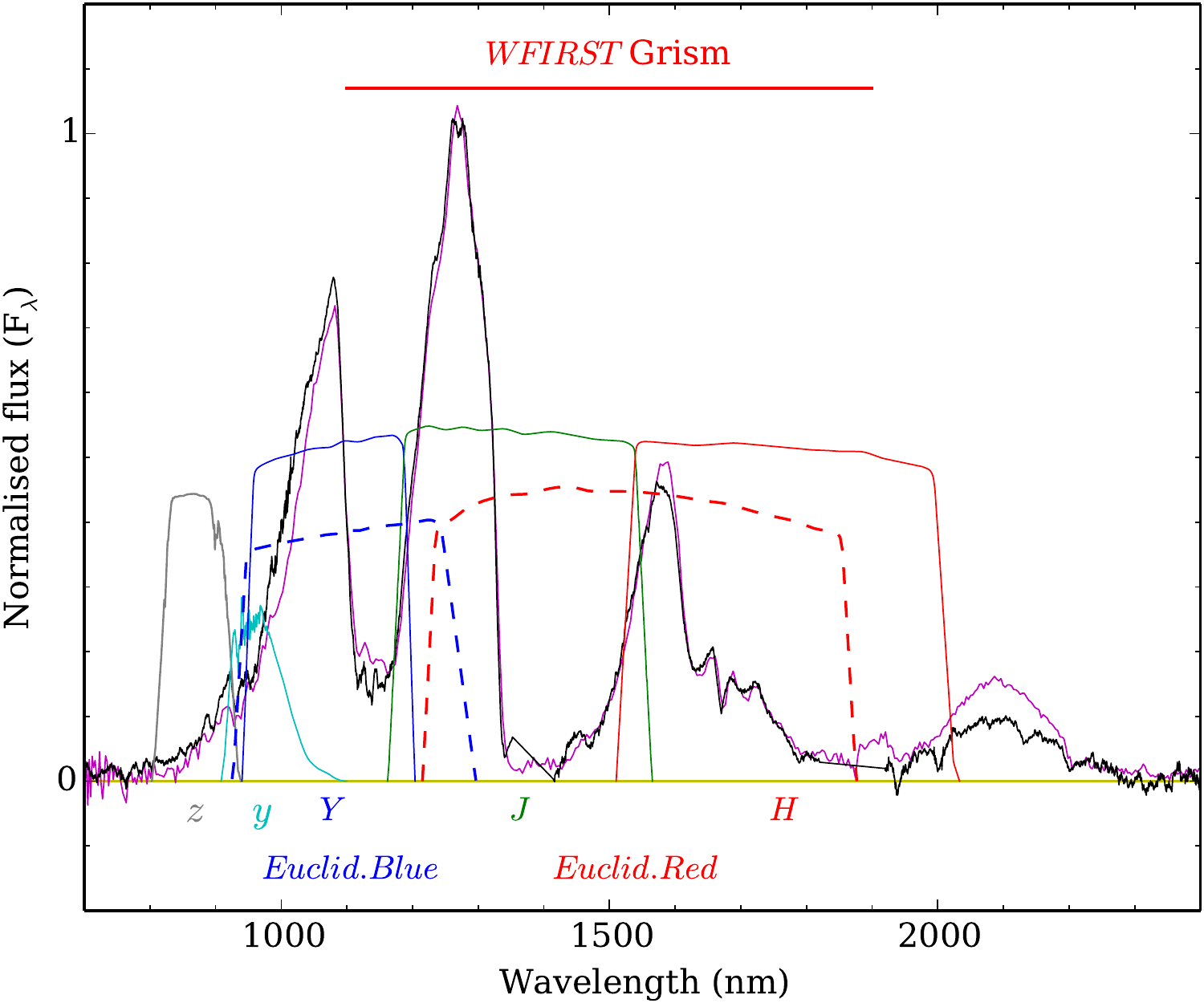}
\caption{Transmission curves of  {\sl Euclid}'s $Y, J, H$ filters for photometry, blue and red filters for slitless spectroscopy \citep{maci16}, and  LSST's $z, y$ filters \citep{ive08}. NIR spectra of an sdT5.5 subdwarf (HIP 73786B; the black line; this paper) and T5.5 dwarf (2MASS J12314753+0847331; the magenta line; \citealt{bur04b}) are plotted as reference. The probable wavelength range of {\sl WFIRST}'s slitless spectroscopy \citep{gong16} is indicated with a red line on the top.   
}
\label{ffil}
\end{center}
\end{figure}

The {\sl Euclid}'s NISP wide survey will cover the peak flux of T subdwarfs in $Y, J, H$ bands. Fig. \ref{ffil} shows transmission curves of {\sl Euclid}'s filters compared to the NIR spectrum of an sdT5.5 subdwarf (HIP 73786B). The characteristic of a T subdwarf spectrum is that they have excess flux at $Y$ band and suppressed flux at $K$ band than a T dwarf with the same subtype. The limiting magnitude (AB) of {\sl Euclid}'s wide survey is about 24 in $Y, J$, and $H$ bands. However, the discovery capability is depending on the depth of $Y$ band ($Y_{\rm Vega} \leq 23.366$), which is the shallowest band for sdT7 subdwarfs. Therefore, {\sl Euclid}'s wide survey could detect about 2147 thick disc and 184 halo T5+ subdwarfs in $Y, J, H$ bands (see Table \ref{teuclid}). 

{\sl Euclid}'s slitless spectroscopic survey could observe 920--1850 nm wavelength low-resolution spectra of about four thick disc T5+ subdwarfs down to $H_{\rm AB}=19.5$ ($H_{\rm Vega}= 18.121$). Note the spectroscopic survey in $J$-band should be slightly deeper than $H$- and $Y$-band for sdT7 subdwarfs (26 pc in $J$-band and 17 pc in $H$-band). 

The {\sl WFIRST}'s High Latitude Survey (HLS) will observe 2000 deg$^2$ of the sky to a depth of $Y_{\rm AB}=26.7$ ($Y_{\rm Vega}=26.066$), and will have a discovery volume of about six times of {\sl Euclid}'s wide survey. The HLS could detect about 11 931 thick disc and 1023 halo T5+ subdwarfs in $Y, J, H$ bands. The $K$-band flux of T subdwarfs is sensitive to metallicity differential. An additional $K$-band filter for  {\sl WFIRST}'s HLS \citep{stau18} would significantly improve its accuracy in the photometric selection of T subdwarfs. Although the survey depth of T subdwarfs would be much shallower in $K$-band ($K_{\rm Vega} = 23.6$), an sdT7 subdwarf at {\sl WFIRST}'s $K$-band limiting magnitude would have $Y_{\rm Vega} = 23.50$ and $J_{\rm Vega} = 22.54$. However, a robust candidate list of about 611 thick disc and 52 halo T5+ subdwarfs selected with $Y, J, H, K$ photometry provides ideal targets for spectroscopic follow-ups with the next generation telescopes ({\sl JWST}, GMT, E-ELT, and TMT). 

The {\sl WFIRST}'s slitless spectroscopic survey is $\sim$2 mag deeper than the {\sl Euclid}'s, and could observe the 1100--1900 nm wavelength spectra of about nine thick disc and one halo T5+ subdwarfs. {\sl WFIRST}'s spectra may not be wide enough to distinguish T5+ subdwarfs with $-0.6 \la$ [Fe/H] $\la -0.3$. Because it is not covering the metallicity-sensitive wavelengths at 950--1100 and 2000--2200 nm (Fig \ref{ffil}). However, {\sl WFIRST}'s low-resolution spectra possibly could distinguish more metal-poor T subdwarfs, which have redder $J-H$ colour than T dwarfs (Fig. \ref{fsc}), and have different $H$-band spectra from T dwarfs.

\begin{table*}
 \centering
  \caption[]{T5--9 subdwarf discovery capability of {\sl Euclid} and {\sl WFIRST}, {\sl WISE}, LSST, and {\sl CSS-OS} surveys.  Limiting magnitudes ($m_{\rm limit}$) in AB system are converted to Vega system according to table 7 of \citet{hew06}. Note the actual discovery number of halo T subdwarfs could be lower (see the last paragraph of Section \ref{sfss}).}
\label{teuclid}
  \begin{tabular}{l c c c c c c c r r r}
\hline
Name & Survey & Filter & Coverage &  $m_{\rm limit}$   &  $m_{\rm limit}$   & $M_{\rm sdT7.5}$   & $M_{\rm sdT7}$ & $d_{\rm sdT7}$ & Thick disc &  Halo \\
 & & & (deg$^2$) & (AB) & (Vega) & (AB) & (Vega) & (pc) & & \\
\hline
{\sl Euclid} & Wide & slitless & 15000 &  $H$ = 19.5 &  18.121 & --- & 16.95 & 17 & 4 & 0 \\
{\sl Euclid} & Wide & $Y$ & 15000 &  24.0 &  23.366 & --- & 17.41 & 155 & 3174 & 272 \\
{\sl Euclid} & Wide & $J$ & 15000 &  24.0 &  23.062 & --- & 16.52 & 203 & 7151 & 613 \\
{\sl Euclid} & Wide & $H$ & 15000 &  24.0 &  22.621 & --- & 16.95 & 136 & 2147 & 184 \\
\hline
{\sl WFIRST} & HLS &  slitless & 2000 &  $H$ = 21.5 &  20.121 & --- & 16.95 & 43 & 9 & 1 \\
{\sl WFIRST} & HLS &  $Y$ & 2000 &  26.7 &  26.066 & --- & 17.41 & 538 & 17640 & 1512 \\
{\sl WFIRST} & HLS &  $J$ & 2000 &  26.9 &  25.962 & --- & 16.52 & 773 & 52395 & 4491 \\
{\sl WFIRST} & HLS &  $H$ & 2000 &  26.7 &  25.321 & --- & 16.95 & 472 & 11931 & 1023 \\
{\sl WFIRST} & HLS &  ($K_s$) & 2000 &  25.5 &  23.600 & --- & 17.38 & 175 & 611 & 52 \\
\hline
{\sl WISE} & {\sl AllWISE} & $W1$ & all sky & --- &  17.9 & --- & 15.90 & 25 & 37 & 3 \\
{\sl WISE} & {\sl AllWISE} & $W2$ & all sky & --- &  16.4 & --- & 13.17 & 44 & 203 & 17 \\
{\sl WISE} & {\sl CatWISE} & $W1$ & all sky & --- &  18.55 & --- & 15.90 & 34 & 91 & 8 \\
{\sl WISE} & {\sl CatWISE} & $W2$ & all sky & --- &  17.05 & --- & 13.17 & 60 & 499 & 43 \\
\hline
LSST & Single-visit  & $z$ & 18000 &  23.3 &  --- & 21.78 & --- & 20 & 8 & 1 \\
LSST & Single-visit & $y$ & 18000 &  22.1 & --- & 19.94 & --- & 27 & 20 & 2 \\
LSST & Coadded &  $z$ & 18000 &  26.1 & --- & 21.78 & --- & 73 & 398 & 34 \\
LSST & Coadded & $y$ & 18000 &  24.9 & --- & 19.94 & --- & 98 & 965 & 83 \\
\hline
{\sl CSS-OS} & Wide &$z$ & 17500 &  25.3 & --- & 21.78 & --- & 51 & 128 & 11 \\
{\sl CSS-OS} & Wide &$y$ & 17500 &  24.7 & --- & 19.94 & --- & 90 & 711 & 61 \\
\hline
\end{tabular}
\end{table*}

\begin{figure}
\begin{center}
  \includegraphics[width=\columnwidth]{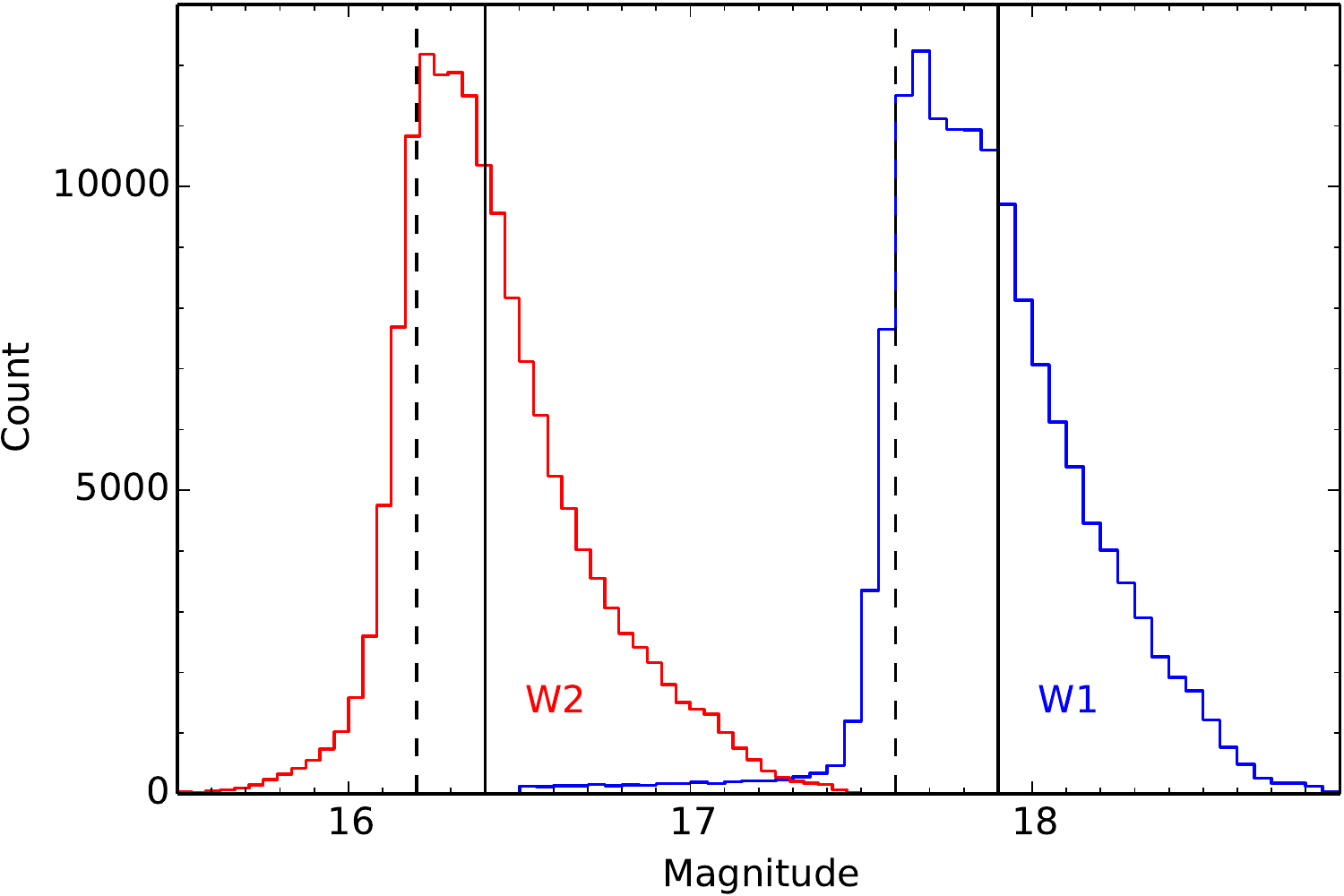}
\caption{{\sl AllWISE} $W1$ and $W2$ magnitude distributions of sources with SNR = 5 selected from half area of the sky ($|b| > 30\degr$). }
\label{fw12}
\end{center}
\end{figure}

Most of the known T5+ subdwarfs were identified with the  {\sl WISE} and {\sl AllWISE} survey (Table \ref{tkmptd}). The survey depth of {\sl AllWISE} varies on the sky. To estimate the mean $W1$ limiting magnitudes of {\sl AllWISE}, we selected a sample of 143 665 sources in {\sl AllWISE} from half area of the sky ($|b| > 30\degr$). We required $0.2 < W1-W2 < 0.4$, $W1 > 16.5$, and SNR$_{W1}$ = 5. The $W1$ distribution of the sample has a peak around $17.6 < W1 < 17.9$, a steep decline between $17.9 < W1 < 18.7$, and a tail extended to $W1 = 18.9$ (Fig. \ref{fw12}). To estimate the mean limiting magnitudes of $W2$, we selected a sample of 157 408 sources with $|b| > 30\degr$, $0.2 < W1-W2 < 0.26$, $W1 > 15.5$, and SNR$_{W2}$ = 5. The $W2$ distribution of the sample has a peak around $16.2 < W2 < 16.4$, a steep decline between $16.4 < W2 < 17.3$, and a tail extended to $W1 = 17.6$ (Fig. \ref{fw12}). We used $W1 < 17.9$ and $W2 < 16.4$ as the mean limiting magnitudes of {\sl AllWISE} in our calculation. 

The {\sl CatWISE} is combining data from {\sl WISE} and {\sl NEOWISE} surveys to produce a $W1$ and $W2$ photometric catalogue about 0.65 mag deeper than {\sl AllWISE}. A small fraction of T subdwarf flux could be detected through $z$ and $y$ filters of the LSST and {\sl CSS-OS} surveys (see Fig. \ref{ffil}). The numbers of thick disc and halo T5--9 subdwarfs that could be detected in each band of {\sl AllWISE}, {\sl CatWISE}, LSST, and {\sl CSS-OS} are listed in Table \ref{teuclid}. 

The selection of T subdwarf candidates normally requires multiband photometry to separate them from other populations. The more bands are applied in the selection, the more efficient it is. Therefore, the discovery capability of a survey is mainly depending on its shallowest band. These shallowest bands are $Y$ band for  {\sl Euclid} and {\sl WFIRST}, $W1$ band for {\sl WISE}, and $z$ band for LSST and {\sl CSS-OS}. However,  $W2$ of {\sl WISE} and $y$ band of LSST or {\sl CSS-OS} could be combined with {\sl Euclid} or {\sl WFIRST} to improve the accuracy in the selection of T subdwarf candidates.

There are 41 known T subdwarfs in Table \ref{tkmptd}. A total 25 were first discovered by {\sl WISE}. In total, 35 of these 41 T subdwarfs were detected in {\sl AllWISE}. A few known T dwarfs might also be metal-poor, but do not have $K$-band spectra in the literature to confirm. {\sl AllWISE} could detect about 40 T subdwarfs in $W1$ band according to our calculation (Table \ref{teuclid}), which is consistent with the number of {\sl AllWISE} $W1$-band detected T subdwarfs in Table \ref{tkmptd}. {\sl AllWISE} could detect about 220 T subdwarfs in $W2$ band, but T subdwarf candidate selection often requires detection in both $W1$ and $W2$ bands. 

The numbers of halo T subdwarfs listed in Table \ref{teuclid} could be overestimated. The discovery volume in our calculation is based on the absolute magnitudes of known T subdwarfs that are mainly sdT subclass and belong to the thick disc population. Fig. \ref{fsptm} shows that the sdT subclass have fainter absolute magnitudes than T dwarfs. Comparably, the esdT and usdT subclasses (halo population) probably have fainter absolute magnitudes than sdT subclass. Although the $T_{\rm eff}$ reduction of T subdwarfs from 0.1 to 0.01 and 0.001 $Z_{\sun}$ is smaller than that from 1 to 0.1 $Z_{\sun}$ according to Fig. \ref{fiso}, the discovery volume (and number) of halo T subdwarfs would be reduced by half if their absolute magnitude is half magnitude fainter than sdT subclass. This may not affect the $W2$ detection, as T dwarfs and sdT subdwarfs have similar $M_{W2}$. Secondly, halo D-BDs have cooler $T_{\rm eff}$ than that of thick disc, thus the fraction of halo D-BDs in the $T_{\rm eff}$ range of Y type is higher than that for thick disc D-BDs. Therefore, the T/Y-type ratio for halo D-BDs (esdT/esdY and usdT/esdY subclasses) is lower than that for thick disc D-BDs (sdT/sdY subclass). 

\section{Summary and Conclusions}
\label{ssum}
We presented 15 new T dwarfs discovered by UKIDSS LAS, VISTA VHS and VIKING surveys. Spectroscopy follow-ups were carried out with GTC/OSIRIS and VLT/X-shooter. One of these objects (UL0021) is mildly metal-poor. One of these objects (UL0008) is likely an unresolved T6+T8 BD binary. We also presented a new optical to NIR X-shooter spectrum of a previous known object \citep[HIP 73786B;][]{murr11}, and re-classfied it as an sdT5.5 subdwarf.  

Nuclear burning transition zones defined by unsteady hydrogen fusion, and incomplete lithium and deuterium fusion exist among BD population. However, their impact on BD population properties are different. The hydrogen burning transition zone is also the STZ that separates stars, T-BDs, and D-BDs. spectral-type and $T_{\rm eff}$ range of T-BDs in the STZ are stretched between stars and D-BDs. Lithium depletion test can be used to identify D-BDs with mass between 0.025 and 0.05 M$_{\sun}$ and age of $\la$2 Gyr. The deuterium burning transition zone is around 0.01--0.014 M$_{\sun}$, which may not possible to reveal by direct observations. However, it is used to define the upper mass boundary (0.01 M$_{\sun}$) of P-BD or gaseous exoplanets.

We extensively studied spectral-type--infrared colour correlation, spectral-type--absolute magnitude correlation, colour--colour plot, and HR diagrams of T5+ subdwarfs. These observational properties of known L subdwarfs are also discussed for comparison. 

Metal-poor D-BDs have become T5+ and Y subdwarfs after over 8 Gyr of evolution. Known T subdwarfs were occasionally discovered in searches for T dwarfs. T dwarfs have condensed atmospheres, the differences of spectral features and photometric colours between dwarfs and subdwarfs caused by metallicity differential are smaller than that for L dwarfs that have cloudy atmospheres (fig. 17; \citetalias{zha18b}). Therefore, the variation of infrared photometric colours caused by metallicity differential of T subdwarfs is less significant than that for L dwarfs. 

T subdwarfs are faint and have larger uncertainty on their photometry, thus have larger scatter on infrared colour--colour plots. Therefore, the photometric selection of T subdwarf candidates with infrared sky survey is usually not efficient. Proper motion or tangential velocity information could significantly improve the selection of T subdwarfs, which are kinematically associated with the Galactic thick disc and halo. However, existing infrared proper motion catalogues either have relatively small sky coverage or are not deep enough for T subdwarfs. 

The future {\sl Euclid}, {\sl WFIRST}, {\sl CatWISE}, LSST, and {\sl CSS-OS} surveys could detect thousands of T subdwarfs of the thick disc and halo. However, it also requires multiband/epoch detections to improve the accuracy of candidate selection. A combined use of these surveys would be very useful in the identification of T subdwarfs.

\section*{Acknowledgements}
Based on observations collected at the European Organisation for Astronomical Research in the Southern Hemisphere under ESO programmes 096.C-0974 and 092.C-0229. Based on observations made with the Gran Telescopio Canarias (GTC), installed in the Spanish Observatorio del Roque de los Muchachos of the Instituto de Astrof{\'i}sica de Canarias, in the island of La Palma. 

This work is based in part on data obtained as part of the UKIRT Infrared Deep Sky Survey. 
The UKIDSS project is defined in \citet{law07}. UKIDSS uses the UKIRT Wide Field Camera \citep[WFCAM;][]{casa07}. The photometric system is described in \citet{hew06}, and the calibration is described in \citet{hodg09}. The pipeline processing and science archive are described in \citet{irwi04} and \citet{hamb08}.
Based on observations obtained as part of the VISTA Hemisphere Survey, ESO Progam, 179.A-2010 (PI: McMahon) and the VIKING survey from VISTA at the ESO Paranal Observatory, programme ID 179.A-2004. Data processing has been contributed by the VISTA Data Flow System at CASU, Cambridge and WFAU, Edinburgh. The VISTA Data Flow System pipeline processing and science archive are described in  \citet{irwi04}, \citet{hamb08} and \citet{cros12}. 

This publication makes use of data products from the {\sl Wide-field Infrared Survey Explorer}, which is a joint project of the University of California, Los Angeles, and the Jet Propulsion Laboratory/California Institute of Technology, funded by the National Aeronautics and Space Administration.
Funding for the SDSS and SDSS-II has been provided by the Alfred P. Sloan Foundation, the Participating Institutions, the National Science Foundation, the U.S. Department of Energy, the National Aeronautics and Space Administration, the Japanese Monbukagakusho, the Max Planck Society, and the Higher Education Funding Council for England. The SDSS Web Site is http://www.sdss.org/. Funding for SDSS-III has been provided by the Alfred P. Sloan Foundation, the Participating Institutions, the National Science Foundation, and the U.S. Department of Energy Office of Science. The SDSS-III web site is http://www.sdss3.org/. 

This research has made use of the VizieR catalogue access tool, CDS, Strasbourg, France. 
Research has benefited from the M, L, and T dwarf compendium housed at DwarfArchives.org and maintained by Chris Gelino, Davy Kirkpatrick, and Adam Burgasser. This research has benefited from the SpeX Prism Spectral Libraries, maintained by Adam Burgasser at http://www.browndwarfs.org/spexprism. This publication makes use of VOSA, developed under the Spanish Virtual Observatory project supported from the Spanish MICINN through grant AyA2008-02156. 

\bibliographystyle{mnras}
\bibliography{primeval6} 

\label{lastpage}
\end{document}